%% file: ckmpp.tex
\documentclass[12pt]{article}

\usepackage[makeroom]{cancel}
\usepackage[normalem]{ulem}
\usepackage{amsmath,amssymb}
\usepackage{slashed}
\usepackage{xcolor}
\usepackage{graphicx}
\usepackage{cite}
\usepackage{pdflscape}
\usepackage{multirow}
\usepackage[thinlines]{easytable}
\usepackage{url}
\usepackage[utf8]{inputenc}

\usepackage{longtable}
\usepackage{booktabs}
\newcommand{\eq}[1]{\begin{equation} #1 \end{equation}}
\newcommand{\eqa}[1]{\begin{eqnarray} #1 \end{eqnarray}}
\newcommand{\eV}{\,{\rm eV}}
\newcommand{\MeV}{\,{\rm MeV}}
\newcommand{\GeV}{\,{\rm GeV}}
\newcommand{\Eq}[1]{Eq.~(\ref{#1})}
\newcommand{\Ref}[1]{Ref.~\cite{#1}}
\newcommand{\Sec}[1]{Section~\ref{#1}}
\newcommand{\App}[1]{Appendix~\ref{#1}}
\newcommand{\Tab}[1]{Table~\ref{#1}}
\newcommand{\B}{{\cal B}}
\newcommand{\av}[1]{\langle #1 \rangle}

\input{shortcuts.tex}

\begin{document}

\allowdisplaybreaks

\begin{titlepage}

\vspace*{-2cm}
\begin{flushright}
LPT Orsay 18-92 \\
CERN-TH-2018-276\\
TUM-HEP-1178/18\\
MIT-CTP/5081
\end{flushright}

\begin{center}
\vspace{1.2cm}
{\LARGE \bf
The CKM parameters in the SMEFT }
\vspace{1.4cm}

\renewcommand{\thefootnote}{\fnsymbol{footnote}}
{S\'{e}bastien Descotes-Genon$^a$, Adam~Falkowski$^a$, Marco~Fedele$^{b}$,\\[2mm]
Mart\'{i}n~Gonz\'{a}lez-Alonso$^{c}$ and Javier~Virto$^{d,e}$}
\renewcommand{\thefootnote}{\arabic{footnote}}
\setcounter{footnote}{0}

\vspace*{.8cm}
\centerline{${}^a$\it Laboratoire de Physique Th\'{e}orique (UMR8627), CNRS, Univ. Paris-Sud,}
\centerline{\it Universit\'{e} Paris-Saclay, 91405 Orsay, France}\vspace{1.3mm}
\centerline{${}^b$\it Dept. de F\'isica Qu\`antica i Astrof\'isica, Institut de Ci\`encies del Cosmos (ICCUB)}
\centerline{\it Universitat de Barcelona, Mart\'i Franqu\`es 1, E08028 Barcelona, Spain}\vspace{1.3mm}
\centerline{${}^c$\it Theoretical Physics Department, CERN, 1211 Geneva 23, Switzerland}\vspace{1.3mm}
\centerline{${}^d$\it Physics Department T31, Technische Universit\"at M\"unchen,}
\centerline{\it James Franck-Strasse 1, D-85748 Garching, Germany}\vspace{1.3mm}
\centerline{${}^e$\it Center for Theoretical Physics, Massachusetts Institute of Technology,}
\centerline{\it 77 Massachusetts Av., Cambridge, MA 02139, USA}

\vspace*{.2cm}

\end{center}

\vspace*{10mm}
\begin{abstract}\noindent\normalsize
The extraction of the Cabibbo-Kobayashi-Maskawa (CKM) matrix from flavour observables can be affected by physics beyond the Standard Model (SM). 
We provide a general roadmap to take this into account, which we apply to the case of the Standard Model Effective Field Theory (SMEFT).
We choose a set of four input observables that determine the four Wolfenstein parameters, and discuss how the effects of dimension-six operators can be included in their definition.
We provide numerical values and confidence intervals for the CKM parameters, and compare them with the results of CKM fits obtained in the SM context.  
Our approach allows one to perform general SMEFT analyses in a consistent fashion, independently of any assumptions about the way new physics affects flavour observables.      
We discuss a few examples illustrating how our approach can be implemented in practice.

\end{abstract}

\end{titlepage}
\newpage 

\renewcommand{\theequation}{\arabic{section}.\arabic{equation}} 

\setcounter{tocdepth}{2}
\tableofcontents


\section{Introduction}
\label{sec:intro}
\setcounter{equation}{0}
Quark flavour-changing transitions have been used to probe high-energy scales since several decades, with many successes along the way, such as predicting the existence and properties of the charm and top quarks. 
They are particularly well suited for this purpose in the context of the Standard Model (SM), where flavour transitions are controlled by the unitary Cabibbo-Kobayashi-Maskawa (CKM) matrix defined by only 4 real parameters. 
Over the last decades, the improvement in measurements has been matched by theoretical progress in computing accurately SM contributions, from the high-energy side (electroweak and perturbative QCD contributions) but also from the low-energy side (hadronic matrix elements involving QCD in the non-perturbative regime).
Global fits to the wealth of experimental data on flavour transitions  show an overall excellent agreement with the SM picture, leading also to a precise determination of the CKM parameters~\cite{Koppenburg:2017mad}.

Such a precise determination of the CKM parameters is essential for precise predictions of many flavour observables, used to set bounds on beyond-the-SM (BSM) physics.
Some of these bounds are among the most stringent BSM constraints available,
and can be translated to lower bounds on the BSM scale much above the reach of
present and near-future colliders.
However, these bounds must be extracted with great care, because the presence of new physics may well invalidate the assumptions implicit in the extraction of the CKM parameters in a SM-based analysis. The goal of this article is to address, in a systematic and model-independent way, how new physics (NP) will affect the global CKM fit and what is the best way to fix the full CKM matrix in a generic BSM context.

Direct searches for new heavy particles at the LHC have not been conclusive, which suggests a significant gap between the electroweak (EW) and BSM scales. In this context, the Standard Model Effective Field Theory (SMEFT)~\cite{Leung:1984ni,Buchmuller:1985jz} represents an 
appropriate theoretical framework to analyse flavour data.
Such data  analysis is done once and for all in the SMEFT and the subsequent correlated bounds on the relevant Wilson coefficients can be applied to a plethora of NP models in a simple way. 
This is true whether one is addressing anomalies in the data, or just setting bounds on BSM physics. 
Moreover, the SMEFT allows one to account for non-trivial correlations between different classes of observables,  such as quark-flavour transitions, leptonic processes, and EW precision measurements. 
In addition it embeds resummations that are needed to tame large logarithms in the presence of large scale hierarchies through Renormalisation Group Evolutions (RGEs).

The SMEFT is the effective theory of any fundamental theory that contains the SM supplemented by
a set of heavy particles with masses $M \sim \Lambda  \gg m_Z$, and in which the EW symmetry is linearly realised. 
Technically speaking, it extends the SM Lagrangian with higher-dimensional operators built from SM fields, with the assumption that these operators also obey the gauge symmetry of the SM.
The leading NP effects are typically encoded in the Wilson coefficients of operators of dimension six, and thus the primary goal is to establish confidence intervals for these parameters.
However, it is important to realise that the free parameters of the SMEFT are not only the Wilson coefficients,
but also the parameters already present in the SM Lagrangian (or the ``SM parameters'' with a slight abuse of language):
the gauge and Yukawa couplings, the Higgs mass and its vacuum expectation value (VEV).  
In a consistent analysis one needs to take into account the presence of the NP contributions affecting the input observables from which the SM parameters are extracted. 

The issue of NP ``contamination" has been carefully studied in the context of EW precision observables (see e.g. \Ref{Wells:2005vk}), where the relevant SM parameters are the $SU(2)_L \times U(1)_Y$ gauge couplings $g_L,g_Y$, and the Higgs VEV $v$. 
In the SM their numerical values are  determined from three very precisely measured experimental inputs (see e.g.~\Ref{pdg18}): the fine-structure constant $\alpha_{\rm em}(0)$ (from the Rydberg constant), 
the $Z$-boson mass (from the $Z$ lineshape in LEP-1), and the Fermi constant $G_F$ (from the muon lifetime). 
In the SMEFT, the relation between these observables and $g_L$, $g_Y$, and $v$ is affected by dimension-six operators, and a straightforward extraction of the SM parameters is thus not possible. 
For example, the Fermi constant is given by 
$G_F = (\sqrt 2 v^2)^{-1} 
\left ( 1 + \delta G_F/G_F \right )$ where 
$\delta G_F$ is a linear combination of several dimension-six Wilson coefficients (see \sref{tildev} for details).  
One option would be to perform a global fit simultaneously to the Wilson coefficients and the SM parameters. 
A more convenient and practical approach, however, consists in absorbing 
$\delta G_F$ into a redefinition of the SM parameters, in a procedure akin to the renormalisation of the SM at one loop. 
Namely, one can define the ``tilde VEV" parameter $\tilde v = v \left( 1+ \delta  G_F/G_F \right)^{-1/2}$ such that it relates to the Fermi constant in the same way as the Higgs VEV in the SM: $G_F = (\sqrt 2 \tilde v^2)^{-1}$, and has a well-defined numerical value, $\tilde v = 246.21965(6)$~ GeV.
Although $\delta G_F$ has been redefined away here, 
it does not disappear.
Instead, NP corrections proportional to $\delta G_F$ emerge in SMEFT predictions of other electroweak observables (such as  
the $W$ boson mass) that  depend on the Higgs VEV in the SM limit, once $v$ is traded for $\tilde v$. 
This kind of approach was followed in some previous global analyses of EW precision observables within the SMEFT, see e.g.~\Ref{Han:2004az,Falkowski:2014tna,Berthier:2015gja,Falkowski:2017pss,Aebischer:2018bkb}.

We want to
develop  an analogous approach to  deal with the CKM parameters in the SMEFT consistently.
This task is much more challenging than in the EW sector where a small set of precisely measured and theoretically clean input observables can be distinguished.  
On the contrary, SM CKM fits rely on many distinct observables, often measured in elaborate experimental set-ups and displaying a complicated dependence on non-perturbative hadronic inputs. These require more involved theoretical approaches, sometimes relying on assumptions only justified within the SM.  For this reason the current global CKM fits developed within the SM cannot be used without a careful adaptation  in a general BSM framework such as the SMEFT. 
Perhaps for these reasons there is, in fact, no complete  SMEFT analysis of flavour data available in the literature to this day.\footnote{%
Consistent analyses of smaller flavour sub-sectors do exist, such as e.g. Refs.~\cite{Gonzalez-Alonso:2016etj,Gonzalez-Alonso:2018omy,Cirigliano:2018dyk}, which study semileptonic light quark transitions.  These processes only involve the Wolfenstein parameter $\lambda$, which is treated as a floating parameter in these references. The non-trivial extension of this approach to the full flavour sector is the subject of this work.} 
Model-independent analyses available in the literature (and also many model-dependent ones) involve different classes of assumptions. A usual approach, motivated by setups with a very high NP scale and an arbitrary flavour structure, is to use $\Delta F=2$ processes to extract NP bounds\cite{Charles:2004jd,Bona:2007vi,Isidori:2010kg,Lenz:2010gu,Lenz:2012az,Charles:2013aka}, with the implicit or explicit assumption that the extraction of the CKM parameters themselves (which are used to calculate the SM prediction of these $\Delta F=2$ processes) is not affected by NP~\cite{Isidori:2010kg}. 
Moreover, it is sometimes assumed  that the ``PDG values" of the CKM parameters (obtained from the global SM CKM fits) can still be used in such BSM setups. Such assumptions are highly non-trivial and they greatly reduce the model-independent nature of these studies.
The goal of this work is to provide the necessary results so that such assumptions are not needed.

In this article we propose a consistent framework to use the CKM parameters in the SMEFT.   
To this end, we will select a number of input observables, and identify specific combinations of the CKM parameters and Wilson coefficients that are determined by these observables.
These combinations will define the ``{\em tilde Wolfenstein parameters}" which, in analogy to the tilde VEV, can be used  to predict numerical values and NP dependence of other flavour observables.  
The outline of the article is the following.
\sref{theory} introduces the theory framework and notation.
In~\Sec{sec:Strategy} we describe our strategy to extract the
CKM parameters in the general context of the SMEFT, and justify our choice of the input observables.
In~\Sec{sec:Analysis} we extract numerical values  for the tilde Wolfenstein parameters from the input observables, and give the necessary formulas to apply our results in phenomenological SMEFT applications. 
In~\sref{applications} with discuss some examples of  applications of our formalism.
\sref{conclusions} contains our conclusions and some future perspectives.


\section{Theoretical framework}
\label{sec:theory}
\setcounter{equation}{0} 

\subsection{Fermion masses and CKM matrix beyond the SM}

We assume that there is a hierarchy between the EW and NP scales ($\mu_{\rm EW}\ll \Lambda_{\rm NP}$), 
and that EW symmetry breaking is linearly realised.
In that case the physics at the EW scale is described by the SMEFT~\cite{Buchmuller:1985jz,Leung:1984ni}:
\eq{
\cL_{\rm SMEFT} = \cL_{\rm SM} + \cL_{D>4} = \cL_{\rm SM} + \sum_i C_i\, Q_i^{(6)} + \cdots \ ,
}
where $Q_i^{(6)}$ and $C_i$ are respectively the dimension-six effective operators and their Wilson coefficients, and
the dots include operators that violate lepton or baryon number and operators of dimension larger than six, which we will not consider.  We will use the Warsaw basis and the notation and conventions in~\Ref{Grzadkowski:2010es} except for the fact that we include $1/\Lambda^2$ in the coefficients $C_i$.

In the broken phase, the Lagrangian at the EW scale contains the fermion mass terms:
\eq{
\cL_{m_\psi} = - \sum_{\psi=u,d,e}
\overline\psi_{R,i} \,[ M_\psi ]_{ij}\, \psi_{L,j} + {\rm h.c.} 
\ ,
}
where the mass matrices include contributions from the SM Yukawa couplings as well as from dimension-six operators:
\eq{
\label{eq:mass}
M_\psi
= - \frac{v}{\sqrt2}  \Big[\Gamma^\dagger_\psi - \frac{v^2}2 C^\dagger_{\psi H}\Big]
\ ,
}
where $v$ denotes the VEV of the Higgs doublet in the presence of dimension-six operators. 

It is always possible to define the Lagrangian in a ``weak" basis for the fermion fields where the mass matrices are given by
\eq{
M_e = {\rm diag}(m_e,m_\mu,m_\tau), \  M_u = {\rm diag}(m_u,m_c,m_t), \  M_d = {\rm diag}(m_d,m_s,m_b) \cdot V^\dagger\ ,
\label{MassMats}
}
with $V$ a unitary matrix. We will adopt this convention, in line with Refs.~\cite{Alonso:2013hga,Jenkins:2017jig}.
Thus, the right- and left-handed lepton and up-quark fields, as well as the right-handed down-quark fields are the same in the
weak and mass eigenstate bases: $e_{L,1} = e_L$, $e_{R,2}=\mu_R$, $u_{L,2}=c_L$, $d_{R,3}=b_R$, etc., while the translation from weak to
mass eigenstate flavour indices for the left-handed down quarks is given by the $V$ matrix:
\eq{
d_{L,i} = V_{ix} \,d_{L,x} = V_{id} \,d_L + V_{is} \,s_L + V_{ib} \,b_L\ , \qquad i=1,2,3\ .
\label{ditodx}
}
Formally, $V_{ix}$ has a weak index $i=\{1,2,3\}$ and a mass-eigenstate index $x=\{d,s,b\}$~\cite{Jenkins:2017jig}.
Since in our convention the weak and mass bases for up-type quarks are the same, it holds that
$V_{1x} = V_{ux}$, $V_{2x} = V_{cx}$ and $V_{3x} = V_{tx}$, and from now on we can use $V_{rx}$ with both mass-eigenstates indices $r={u,c,t}$ and $x={d,s,b}$.
In this article we use the Wolfenstein parameterization for $V$:
\eqa{
\label{eq:CKM_vckm}
V &=&
\left(
\begin{array}{ccc}
V_{ud} & V_{us} & V_{ub} \\
V_{cd} & V_{cs} & V_{cb} \\
V_{td} & V_{ts} & V_{tb} \\
\end{array}
\right)
\\[3mm]
&=& 
\small
\left (\ba{ccc}
1 - \frac12 \lambda^2 - \frac18 \lambda^4  & \lambda &  A \lambda^3 (1 + \frac12 \lambda^2) (\bar\rho - i \bar\eta)  \\[1mm]
-\lambda+ A^2\lambda^5 (\frac12-\bar\rho-i\bar\eta) & 1 - \frac12 \lambda^2 - \frac18 \lambda^4 (1 + 4  A^2)  &  A \lambda^2 \\[1mm]
 A \lambda^3(1 - \bar\rho - i \bar\eta) & -  A \lambda^2  +  A\lambda^4    (\frac12 - \bar\rho - i \bar\eta )   & 1  - \frac12 A^2\lambda^4  
\ea \right ) + \cO(\lambda^6)\ .
\nonumber
}
We refer to the unitary matrix $V$ as {\em the CKM matrix}. 
Its definition is affected by the presence of certain dimension-six operators, {\it cf.}~\eref{mass}. Moreover, and contrary to the SM, the flavour structure of charged currents is not uniquely determined by the CKM matrix, but is also affected by the presence of dimension-six operators with generic flavour structure.\footnote{%
A tacit assumption throughout this article is that the numerical values of the Wolfenstein parameters in the SMEFT are not far from the ones determined in the SM context; in particular that $\lambda$ is small enough to serve as an expansion parameter in \eref{CKM_vckm}.}  
In the following we discuss the consistent extraction of $V$ from flavour observables within the general context of the SMEFT.

\subsection{Effective theory below the EW scale}

While it is possible 
that, in the future, precision high-energy measurements at the EW scale might be used to extract the parameters of the CKM matrix
(see e.g.~\cite{Harrison:2018bqi}), low-energy flavour-violating observables remain currently the best window to CKM physics. 
These  observables
are calculated in an effective theory where particles with EW-scale masses have been integrated out~\cite{Buchalla:1995vs, Cirigliano:2009wk,Aebischer:2017gaw,Jenkins:2017jig}. 
Low-energy flavour observables probe directly the Wilson coefficients of the operators in this Low-energy EFT (LEFT) at the appropriate hadronic scale,
which can be related to the SMEFT through RGE together with a matching at the EW scale.

In this article we will use the LEFT basis and notation of Ref.~\cite{Jenkins:2017jig}:
\eq{
\cL_{\rm LEFT} = \cL_{\rm QED+QCD} \,+  \sum_i L_i\, \cO_i^{(5,6)} + \cdots \ ,
}
where we have kept lepton- and baryon-number conserving operators of dimension five and six, $\cO_i^{(5,6)}$, with $L_i$ denoting the respective Wilson coefficients.
For $B$ physics, this EFT includes all quarks and leptons except the top quark~\cite{Aebischer:2017gaw}, while for physics at lower energies one may integrate out additional fields such as the $b$ quark.
Anticipating the relevant observables that will be chosen in~\Sec{sec:CKM} to fix the CKM matrix, 
we focus on the  semileptonic and $\Delta F=2$ operators (mediating $d\to u \mu^- \bar \nu_\mu$, $s\to u \mu^- \bar \nu_\mu$, $b\to u \tau^- \bar \nu_\tau$ transitions 
as well as $B_{d}$ and $B_{s}$ mixings). 
The relevant LEFT operators are collected for convenience in~\Tab{tabops}.

\begin{table}
\centering
\setlength{\tabcolsep}{14pt}
\renewcommand{\arraystretch}{1.5}
\begin{tabular}{@{}ll@{}}
\toprule
Semileptonic  & $\Delta F=2$\\
\hline
$[\cO_{\nu edu}^{V,LL} \big]_{iijk} =  (\bar \nu_{L,i} \gamma^\mu e_{L,i})(\bar d_{L,j} \gamma_\mu u_{L,k})$    
& $[\cO_{dd}^{V,LL}]_{ijij} = (\bar d_{L,i} \gamma^\mu d_{L,j})(\bar d_{L,i} \gamma_\mu d_{L,j})$ \\
$[\cO_{\nu edu}^{V,LR} \big]_{iijk} =  (\bar \nu_{L,i} \gamma^\mu e_{L,i})(\bar d_{R,j} \gamma_\mu u_{R,k})$    
& $[\cO_{dd}^{V,RR}]_{ijij} = (\bar d_{R,i} \gamma^\mu d_{R,j})(\bar d_{R,i} \gamma_\mu d_{R,j})$ \\
$[\cO_{\nu edu}^{S,RR} \big]_{iijk} =  (\bar \nu_{L,i}  e_{R,i})(\bar d_{L,j} u_{R,k})$    
& $[\cO_{dd}^{V1,LR}]_{ijij} = (\bar d_{L,i} \gamma^\mu d_{L,j})(\bar d_{R,i} \gamma_\mu d_{R,j})$ \\
$[\cO_{\nu edu}^{T,RR} \big]_{iijk} =  (\bar \nu_{L,i} \sigma^{\mu\nu} e_{R,i})(\bar d_{L,j} \sigma_{\mu\nu} u_{R,k})$    
& $[\cO_{dd}^{V8,LR}]_{ijij} = (\bar d_{L,i} \gamma^\mu T^a d_{L,j})(\bar d_{R,i} \gamma_\mu T^a d_{R,j})$ \\
$[\cO_{\nu edu}^{S,RL} \big]_{iijk} =  (\bar \nu_{L,i}  e_{R,i})(\bar d_{R,j}  u_{L,k})$    
& $[\cO_{dd}^{S1,RR}]_{ijij} = (\bar d_{L,i} d_{R,j})(\bar d_{L,i}  d_{R,j})$ \\
& $[\cO_{dd}^{S8,RR}]_{ijij} = (\bar d_{L,i} T^a d_{R,j})(\bar d_{L,i}  T^a d_{R,j})$ \\[1mm]
\bottomrule
\end{tabular}
\caption{Operators in the LEFT~\cite{Jenkins:2017jig} relevant for semileptonic  charged-current transitions and $B_{d,s}$ mixing.}
\label{tabops}
\end{table}

The tree-level matching conditions for the full set of Wilson coefficients $L_i$ in terms of the SMEFT Wilson coefficients can be found in~\Ref{Jenkins:2017jig}.
The matching conditions in the  
SM are known to much higher orders (see e.g.~\cite{Buchalla:1995vs,Bobeth:1999mk}),
while some one-loop contributions from non-SM operators are also known~\cite{Aebischer:2015fzz,Bobeth:2017xry,Endo:2018gdn}.
We will consider the state-of-the-art SM matching conditions but
only tree-level matching from dimension-six operators in the SMEFT, as given in~\Ref{Jenkins:2017jig}. A typical matching condition has the structure:
\eq{
L_i(\mu_{\rm EW}) = F^{\rm SM}_i(\vec g, \vec m,\mu_{\rm EW}) + \sum_j F^{(6)}_{ij}(\vec g, \vec m,\mu_{\rm EW})\, C_j(\mu_{\rm EW})
}
where $F^{\rm SM}_i(\vec g, \vec m,\mu_{\rm EW})$ are the SM matching conditions as functions of the set of couplings and masses in the SMEFT
(collectively called $\vec g$ and $\vec m$), and the product $F^{(6)}_{ij}(\vec g, \vec m,\mu_{\rm EW})\, C_j(\mu_{\rm EW})$ denotes the contribution from the dimension-six SMEFT operator $Q_j^{(6)}$.
The relevant expressions for the specific SMEFT operators needed in the analysis of~\Sec{sec:Analysis} are given in~\App{app:LEFTSMEFTmatching}.

The low-energy amplitudes used to compute the processes of interest for the CKM parameters are given by default in terms of LEFT Wilson coefficients at a low, hadronic scale, where non-perturbative matrix elements are computed:
\mbox{$L_i(\mu_b\sim 4.3 \GeV)$} in the case of $B$ physics and \mbox{$L_i(\mu_s\sim 2 \GeV)$} in the case of $K$ physics.
In order to relate the coefficients at these scales with the matching conditions at the EW scale without generating large logarithms one needs to use RGEs. As in the case of the matching conditions, the anomalous dimensions of the SM operators are known to high orders in $\alpha_s$.
For some sets of BSM operators, two- or three-loop anomalous dimensions in QCD are also known, including the operators that will be relevant in \Sec{sec:Analysis}~\cite{Buras:2000if,Gonzalez-Alonso:2017iyc}. 
One-loop anomalous dimensions in QED+QCD are known for the full set of LEFT operators~\cite{Aebischer:2017gaw,Jenkins:2017dyc}, which are implemented in publicly available software tools~\cite{Celis:2017hod,Aebischer:2018bkb}.
The RG evolution, which can be implemented matricially as
\eq{
L_i(\mu_1) = \sum_j \,[\eta(\mu_1,\mu_2)]_{ij}\, L_j(\mu_2)\ , 
}
is included in the relevant formulas given in~\App{app:LEFTrunning}.


\section{Model-independent determination of the CKM matrix}
\label{sec:Strategy}
\setcounter{equation}{0}

\subsection{Basics of the CKM fit in the SM
}
\label{sec:SMcase}

In the SM, the numerical values of the Wolfenstein parameters $W_i \equiv \{\lambda, A, \bar \rho, \bar \eta\}$ are extracted from a global fit to a long list of experimental observables (see e.g.~\cite{Koppenburg:2017mad}) that are accurately measured and whose SM predictions
are well understood.
They can be separated in four broad categories:
\begin{itemize}

\item {\bf Leptonic decays ($\Delta F=1$ branching ratios):} the branching ratios provide information on the modulus of a CKM matrix element, provided that one knows the corresponding decay constant, i.e. the coupling between the axial current and the relevant meson. Currently, the main observables with accurate theoretical and experimental inputs are
\eqa{
&&
\pi\to \mu\nu \,,\ \ 
K\to e\nu \,,\ \ 
K\to \mu\nu \,,\ \ 
\tau \to K\nu \,,\ \ 
\tau \to \pi\nu \,,
\nonumber\\
&&
D\to \mu\nu \,,\ \ 
D_s\to\mu\nu \,,\ \ 
D_s\to\tau\nu \,,\ \ 
B\to \tau\nu\ .
}
    
\item {\bf Semileptonic decays ($\Delta F=1$ branching ratios):} these measurements provide information on the modulus of a CKM matrix element, provided that one knows the corresponding form factors, i.e. the couplings between the vector/axial and scalar/pseudoscalar currents and the relevant mesons. 
Currently, observables with accurate theoretical and experimental inputs are:
\eq{
K\to\pi e \nu \,,\ \ 
D \to \pi e \nu\, , \ \ 
D\to K e\nu \,,\ \ 
B\to\pi e\nu \,,\ \ 
B\to D e\nu \,,\ \ 
B\to D^* e\nu\ .
}
Also in this category we can include the superallowed nuclear transitions, which provide a precise value for $|V_{ud}|$ and rely on a detailed description of nuclei and their weak transitions.
There are also determinations of $|V_{ub}|$ and $|V_{cb}|$ from inclusive $b\to c\ell\nu$ and $b\to u\ell\nu$ transitions, which are however not fully compatible with the determinations from exclusive transitions (hinting at underestimated systematics). We can also include in this category $|V_{us}|$ determinations from inclusive $\tau\to \bar{u}s\bar{\nu}$ decays.

\item {\bf CP-asymmetries ($\Delta F=1$ CP-violating observables):} These measurements allow one to extract CP-violating phases ($\alpha$, $\beta$ and $\gamma$, see~\Ref{Koppenburg:2017mad} for their definition). They typically combine information from different channels or exploit time-dependent asymmetries involving the same hadronic matrix elements, so that the latter can be determined from the data or cancel out in ratios depending only on the CKM elements.
The presence of the same hadronic matrix elements may hinge on SM flavour symmetries (isospin symmetry for the angle $\alpha$), the hierarchy of CKM contributions (neglect of penguins for the angles $\beta$ and $\beta_s$) or the knowledge of hadronic matrix elements from other sources (hadronic $D$ decays for the angle~$\gamma$). The main current processes of interest are:
\begin{align}
&
B\to \pi\pi,\rho\pi,\rho\rho \quad (\text{for } \alpha) \,,\quad 
&&
B\to J/\psi K^{(*)},(c\bar{c}) K \quad (\text{for } \beta) \,,\ 
\nonumber\\
&
B\to D^{(*)}K^{(*)} \quad (\text{for } \gamma) \,,\quad 
&&
B_s\to J/\psi \phi, \psi(2S) \phi \quad  (\text{for } \beta_s)\ .
\label{nonleptonics}
\end{align}

\item {\bf Neutral-meson mixing ($\Delta F=2$ observables):} these measurements deal with the time evolution of the system composed by a neutral-meson flavour state and its CP-conjugate. They measure properties of the transition from one mass eigenstate of a neutral-meson system to the other (difference of masses, CP violation). They rely on the knowledge of the matrix element of the relevant $\Delta F=2$ operator in the SM between both neutral mesons. The main current observables of interest are
\begin{equation}
\label{eq:mixingobservables}
\epsilon_K\ (K\bar{K})\ ,\quad
\Delta M_d\ (B_d\bar{B}_d)\ ,\quad
\Delta M_s\ (B_s\bar{B}_s)\ .
\end{equation}

\end{itemize}
All these measurements show a remarkable agreement with the CKM picture for the quark-flavour transitions embedded in the SM, leading to an accurate determination of the four CKM parameters~\cite{Koppenburg:2017mad}. 
However, these results have to be reassessed  in the presence of NP, as non-SM contributions may not respect the assumptions implicit in their derivation. 
In the remainder of this article we propose an algorithm for extracting the CKM parameters in a general SMEFT framework,  
and discuss how to translate the measurements of flavour observables into constraints on NP 
in a consistent way.

\subsection{Interlude: Higgs VEV in the SMEFT}
\label{sec:tildev}

Before embarking on the extraction of the CKM parameters in the SMEFT, it is  worth recalling how parameters in the EW sector can be defined to illustrate this strategy.   
We take as an example
the Higgs VEV in the SMEFT.
In the SM, $v$ is related to the Fermi constant $G_F$, which in turn can be defined as a coefficient of the 4-fermion interaction between muons, electrons and neutrinos in the effective theory at a scale $\mu \sim m_\mu$:
\beq
\cL_{\rm eff} \supset 
- 2 \sqrt{2} G_F (\bar \nu_\mu \gamma_\alpha  \,\mu_L)   (\bar e_L \gamma^\alpha \nu_e)  + \hc   
\eeq
Integrating out the tree-level $W$ exchange in the SM one finds
$G_F = (\sqrt {2} v^2)^{-1}$. 
Given the experimental value  $G_F = 1.1663787(6) \times 10^{-5}~\gev^{-2}$~\cite{pdg18} precisely measured in muon decay, one can assign the numerical value to the Higgs VEV, $v = 246.21965(6) \GeV$. 
However, this logic is perturbed  if the SMEFT (and not the SM) is the relevant theory at $\mu \gtrsim m_W$. 
In that case one finds that dimension-six operators affect the Fermi constant as\footnote{
Summing over the 4-lepton terms 
$\sum_{ijkl}\big[ C_{\ell\ell} \big]_{ijkl}
\bar \ell_i \gamma_\mu \ell_j \bar \ell_k \gamma^\mu \ell_l
$ in the SMEFT Lagrangian, the Wilson coefficients  $\big[ C_{\ell\ell} \big]_{ijji}$ and  
$\big[ C_{\ell\ell} \big]_{jiij}$ are indistinguishable because they multiply exactly the same operator. 
In the literature one often encounters the convention that the two Wilson coefficients in this pair are equal, or that one of them is zero. 
Our \eref{GFsmeft} is valid regardless of the convention. 
}
\eqa{
\label{eq:GFsmeft}
G_F &=& {1 \over \sqrt 2 v^2} \bigg (1  + \frac{\delta G_F}{G_F} \bigg) , \nonumber\\[2mm] 
\frac{\delta G_F}{G_F} 
&=& v^2\, \Big( 
\big[ C_{H\ell}^{(3)} \big]_{\mu\mu}
+  \big[ C_{H\ell}^{(3)} \big]_{ee}
- {1 \over 2} \big[ C_{\ell\ell} \big]_{\mu ee\mu}
- {1 \over 2} \big[ C_{\ell\ell} \big]_{e\mu\mu e}
\Big)+  \cO(\Lambda^{-4}),
}
where $v$ is the VEV of the Higgs field in the presence of  dimension-six operators, and $C_{\ell\ell}$, $C_{H\ell}^{(3)}$ are Wilson coefficients of the corresponding operators in the Warsaw basis~\cite{Grzadkowski:2010es}. 
At this stage we cannot assign a numerical value to $v$ without knowing the Wilson coefficients.
Instead, it is convenient to define the tilde VEV parameter $\tilde v$ via the relation
\beq
\label{eq:vtilde}
\tilde v = {v \over \sqrt{1 + {\delta G_F}/{G_F}}} = 
v\,\left(1 + \frac{\delta v}{v} \right) \ ,
\qquad 
\frac{\delta v}{v}  =  - {1 \over 2} \frac{\delta G_F}{G_F}   + \co(\Lambda^{-4}).
\eeq
With this definition we recover $G_F = (\sqrt {2} \tilde v^2)^{-1}$,  
and we can assign a numerical value to $\tilde v$, which is equal to that of $v$ in the SM context, $\tilde v = 246.21965(6) \GeV$. 
In fact, this procedure is similar to
the renormalisation of the SM at one loop. Let us however stress
that we are dealing with finite tree-level corrections
in the present situation.

At this point the dependence of the muon decay width on the SMEFT Wilson coefficients has been absorbed into the definition of $\tilde v$, hence this observable alone does not constrain NP.    
However, the physical effect of  $\delta G_F$ is not void. 
Using \eref{vtilde}, we should replace $v$ with $\tilde v$ in the expression for any other EW observable sensitive to the Higgs VEV in the SM limit, 
in order to isolate the SM prediction for that observable. 
This way, $\delta G_F$ will modify the linear combination of Wilson coefficients to which the observable is sensitive:
\eqa{
&& O = O_{\rm SM}(v) + \delta O^{\rm direct}_{\rm NP} = O_{\rm SM}(\tilde{v}) + \delta O^{\rm indirect}_{\rm NP} + \delta O^{\rm direct}_{\rm NP}~,
\nonumber\\[2mm]
&& \delta O^{\rm indirect}_{\rm NP} = \frac{\tilde v}2 \frac{\delta G_F}{G_F} { \partial O_{\rm SM}(\tilde v) \over \partial \tilde v}  + \cO(\Lambda^{-4})\ .
\label{eq:EWPOindirectNP}
}
The ``direct'' contribution comes from the computation using the initial SMEFT parameters, whereas the ``indirect'' part comes from the redefinition of the Higgs VEV.
We remark that the separation between direct and indirect NP effects is 
semantic. Both effects are in general equally large and physical, namely ${\cal O}(\Lambda^{-2})$ in the SMEFT expansion.

\subsection{Strategy for the extraction of the CKM parameters in the SMEFT}
\label{sec:CKM}

We turn to the determination of CKM parameters in the general context of the SMEFT. 
There are two distinct strategies one could envisage here. 
One could aim at performing a global fit to all available flavour observables where not only the dimension-six Wilson coefficients but also the 4 independent CKM parameters are considered free parameters.
Treating the unknown Wilson coefficients as nuisance parameters
would return confidence intervals for the Wolfenstein parameters.  
This is a formidable task, and a much greater challenge than the SM CKM fit considering the number of parameters involved. An illustration of this strategy in the more limited case of NP only in the $\Delta F=2$ sector can be found in Refs.~\cite{Lenz:2010gu,Lenz:2012az,Charles:2013aka}.

In this article
we opt for a simpler strategy. 
We will identify a minimal set of four optimal observables constraining specific combinations of CKM and SMEFT parameters. 
These observables will define the 4 Wolfenstein parameters to which we will assign numerical values and errors. These in turn can be used to predict numerical values of other flavour observables, which can be compared with the experimental values in order to constrain NP.     
Note that our strategy can be embedded in the former one at a later stage, by using the results obtained in this article as ``pseudo-observables" in a global fit.

In line with the discussion in~\sref{tildev}, we will
denote the combinations of Wolfenstein parameters and SMEFT Wilson coefficients extracted from the selected observables by
(denoting $\tilde \rho \equiv \tilde \barrho$ and $\tilde \eta \equiv \tilde \bareta$)
\begin{equation}
\widetilde{W}_j = \{\tilde\lambda, \tilde A, \tilde \rho, \tilde \eta\}\ ,
\end{equation}
with the understanding that in the SM limit $\widetilde{W}_j \to \{\lambda, A, \bar \rho, \bar \eta\}$.
We will often refer to these quantities as {\em tilde Wolfenstein parameters}, or simply {\em tilde parameters}.

In order to determine $\widetilde{W}_j$ 
one should choose a quartet of observables from the pool of observables listed in~\Sec{sec:SMcase}. 
Ideally, we want the input observables to satisfy the following rather constraining conditions: 

\begin{enumerate}
	
\item The set of observables must have a good sensitivity to all the four Wolfenstein parameters;\label{enu:CKM}
	
\item Each observable should be accurately measured and its theoretical prediction (in the general BSM case) should be well understood and non-controversial;\label{enu:obs}
	
\item The general SMEFT expression for the observables should
involve as few SMEFT Wilson coefficients as possible, in order to minimize the number of correlated observables needed in phenomenological applications.
\label{enu:SMEFT}
	
\end{enumerate}
This greatly reduces the available choices from the list in~\Sec{sec:SMcase}. 
Condition~\#\ref{enu:CKM} is obviously mandatory and can be checked in the SM limit. 
Condition~\#\ref{enu:obs} disfavours $b\to c\ell\nu_\ell~(\ell=e,\mu)$ transitions due to the tensions observed between exclusive vs. inclusive determinations~\cite{Koppenburg:2017mad}. \footnote{It has been noted that using the so-called BGL parameterization for the form factors ameliorates this tension~\cite{Bigi:2017njr,Grinstein:2017nlq}. However, a recent analysis performed by BaBar employing the BGL parameterization found again tension between exclusive and inclusive determinations~\cite{Dey:2019bgc}. We therefore prefer not to include the exclusive determination of $V_{cb}$ as an input observable until the issue gets clarified definitely.}
Condition~\#\ref{enu:SMEFT} is needed to select observables that can be used in the general SMEFT framework while depending only on a limited set of theoretical inputs and unknown parameters.

Let us now discuss some classes of the observables from the CKM fit in the SM in~\Sec{sec:SMcase} in more detail.
Observables from non-leptonic decays in~\Eq{nonleptonics} involve a limited set of hadronic matrix elements in the SM, which can be determined or eliminated thanks to additional observables and symmetries. 
Beyond the SM, however, these observables
involve a much wider set of hadronic matrix elements that are currently not known and, in a general SMEFT context, cannot be related to other hadronic quantities through flavour symmetries. A similar issue affects $\epsilon_K$, which can be extracted from $K\to\pi\pi$ decays only under specific assumptions about the weak amplitudes.

Concerning the semileptonic decays such as $K\to \pi \ell\nu$, $D\to K\ell\nu$, or $B\to \pi\ell\nu$,
the rates depend on form factors whose momentum dependence is usually extracted from the measurement of the differential distributions,
which are themselves modified by BSM effects. Thus in order to use this information, a new BSM analysis of both differential distribution and rate is required (see e.g. Ref.~\cite{Gonzalez-Alonso:2016etj}). This is in contrast to the leptonic decays, whose hadronic input is limited to decay constants,
well known from lattice QCD. 
In addition, semileptonic decays are often sensitive to a larger set of BSM operators than leptonic decays,
disfavouring semileptonic decays on the basis of condition~\#\ref{enu:SMEFT}. 
Overall these arguments favor using leptonic as opposed to semileptonic decays as our input observables.

We can now determine
the most appropriate observables for the determination of the CKM parameters.
Concerning observables sensitive (only) to $\lambda$, condition~\#\ref{enu:obs} suggests to disfavour $D$ and $D_s$ meson decays compared to $K$ decays. 
The latter are measured with a better accuracy and thus exhibit better sensitivity to $\lambda$.
One technical complication, however, arises due to the  dependence of 
the leptonic $K$ decays on the decay constant $f_{K^+}$,
as its most recent determinations rely on  the ``experimental'' value of $f_\pi$ from $\pi\to\mu\nu$ to set the reference scale in the lattice QCD calculations~\cite{Aoki:2019cca}. 
This reintroduces an SM assumption (i.e., that the pion leptonic decay is completely dominated by SM contributions) that is not appropriate for a general analysis in the SMEFT setup~\cite{Gonzalez-Alonso:2016etj}.
To avoid this complication, we take the ratio $\Gamma(K\to \mu \bar \nu)$ to $\Gamma(\pi\to \mu \bar \nu)$ as our input observable, as the lattice determinations of $f_{K^+}/f_{\pi^+}$ are free from this
problem (and known with higher accuracy).
Concerning the parameter $A$, we may consider observables sensitive to $V_{ub}$, $V_{cb}$, $V_{td}$, or $V_{ts}$,
while the highest sensitivity to $\bar \rho$ and $\bar \eta$ comes from $V_{ub}$ and $V_{td}$.
All in all, 
the remaining observables satisfying our criteria and sensitive to these three CKM parameters are  $B\to\tau\nu$ (for $V_{ub}$), $\Delta M_d$ (for $V_{td}$),  and $\Delta M_s$ (for $V_{ts}$). 

This leaves us with the following set of input observables that we consider optimal:
\beq
\label{eq:inputobservables}
\boxed{ \Gamma(K\to\mu\nu_\mu)/\Gamma(\pi \to\mu\nu_\mu), \quad \Gamma(B\to\tau\nu_\tau), \quad \Delta M_d, \quad \Delta M_s.}
\eeq
These four observables indeed obey the criteria listed above. 
In \sref{Analysis} we will show that they provide an accurate determination of the four Wolfenstein parameters  $\widetilde{W}_j$ in the generic SMEFT case, with only a moderate loss of accuracy compared the 
SM case. 
One should stress that our choice is not set in stone, and  some variations on the input observables are of course possible, 
similarly to different input schemes used in EW precision physics.
Furthermore, we emphasise that the ``optimal choice" may vary over time.
For example, if the inclusive-vs-exclusive tensions for $b\to c$ or $b\to u$ transitions disappear, or (theoretical or experimental) progress is achieved in some of the flavour transitions that we dismissed, our input observables may need to be appropriately reconsidered.

\bigskip

Summarizing, our approach to constraining NP in the SMEFT using flavour observables adheres to the following algorithm:

\begin{enumerate} 

\item We identify the dependence
of the input observables $O_i^{\rm input}$ in \eref{inputobservables} on the LEFT Wilson coefficients $L_k$ and, given the LEFT-to-SMEFT map, on the Wilson coefficients of dimension-six operators in the SMEFT $C_k$:
\eqa{
\label{eq:Obsm}
O_i^{\rm input} 
= O_{i,{\rm SM}}^{\rm input} (W_j) \big[(1 + f (L_k) \big]
= O_{i,{\rm SM}}^{\rm input} (W_j)  \big[1 + g (C_k ) \big]
~,
}
where we keep the dependence of the $f$ and $g$ functions on the CKM parameters $W_j \equiv \{\lambda, A, \bar \rho, \bar \eta\}$ as implicit.
\item 
We define the parameters $\widetilde{W}_j \equiv \{\tilde\lambda, \tilde A, {\tilde \rho}, {\tilde \eta}\}$ by
\bea
\label{eq:tildewth}
\widetilde{W}_j = W_j \left(1 + \frac{\delta W_j}{W_j} \right)~.
\eea
where $ \delta W_j / W_j$ are functions of the LEFT (or SMEFT) Wilson coefficients. 
They are defined such that the input observables depend explicitly only on $\widetilde W_j$ in a way similar to the SM expression and involving no additional LEFT or SMEFT Wilson coefficients:
\bea
\label{eq:Osm}
O^{\rm input}_i = O_{i,{\rm SM}}^{\rm input} (\widetilde{W}_j)~.
\eea

\item We extract the numerical value of the tilde parameters $\widetilde{W}_j$ using~\eref{Osm} along with the necessary experimental and theoretical inputs (as described above), keeping full track of correlations. These values, their correlated uncertainties, and the  contribution from BSM operators, given in~\eref{tildewth}, will be the main result of this work. 

\item At this point we can translate any other flavour measurement, $O_\alpha$, into a model-independent NP constraint:
\bea
\label{eq:CKMindirectNP}
O_\alpha 
= O_{\alpha,\rm SM}(W_j) + \delta O^{\rm direct}_{\alpha,\rm NP} 
= O_{\alpha,\rm SM}(\widetilde{W}_j) + \delta O^{\rm indirect}_{\alpha,\rm NP} 
+ \delta O^{\rm direct}_{\alpha, \rm NP}~,
\eea
where $\delta O^{\rm direct}_{\alpha,\rm NP}$ stands for a combination of  Wilson coefficients contributing directly to the observable, and the indirect contribution is~\footnote{These expressions are valid at the linear order in the NP contributions.
In general~\eref{CKMindirectNP} contains also higher order terms in $\delta W_i$, or cross terms of order $\delta W_i \times  \delta O^{\rm direct}_{\alpha,\rm NP}$ that should be included if one wants to trace NP effects beyond the leading order.}
\beq
\delta O^{\rm indirect}_{\alpha, \rm NP}  = - {\partial O_{\alpha,\rm SM} \over \partial W_i} \delta W_i \,+\,\cO(\Lambda^{-4}) .
\eeq 
\eref{CKMindirectNP} is the flavour analogue of~\eref{EWPOindirectNP} relevant for EW precision observables.

\end{enumerate} 

Once the tilde Wolfenstein parameters have been determined, it is convenient to introduce the tilde CKM matrix $\widetilde V$. 
Given the SM expression $V(\lambda, A, \bar \rho, \bar \eta)$ 
in \eref{CKM_vckm}, we define  $\widetilde V$ by
\beq
\label{eq:tildeVjk}
\widetilde V \equiv V(\tilde \lambda, \tilde A, \tilde \rho, \tilde \eta). 
\eeq 
The elements of this matrix can be used to calculate the numerical SM predictions for observables depending on the CKM parameters. The NP effects included in them should be taken into account through the method described above. 
The $\widetilde V$ matrix defined above is unitary by construction. 
This does not entail any loss of generality, because we do not define the nine elements of $\widetilde V$ as the elements extracted from nine different measurements (such matrix would not be unitary in the SMEFT). Unitarity is a key and necessary ingredient, since we only have four independent CKM parameters to fix, and thus we only need to ``lose" four measurements (and not nine) to fix them. Any additional observable becomes in this way a NP probe, as it should be.

A last comment is in order concerning the choice of the hadronic inputs related to these observables. Lattice QCD provides a self-consistent theoretical framework to compute these inputs in global CKM analysis, but it still requires 
phenomenological inputs to determine the values of the parameters of the Lagrangian: the quark masses and the strong coupling constant (i.e. the lattice scale in physical units). 
However, as discussed above for $f_K$, the ``experimental'' value of $f_\pi$ from $\pi\to\mu\nu$ is often used to set the scale in the lattice QCD calculations. This reintroduces an SM assumption (i.e., the pion leptonic decay is completely dominated by SM contributions), which will propagate in all dimensionful lattice QCD inputs and which is thus not appropriate for a general analysis in the SMEFT setup.
From this point of view it is thus better to use determinations of the scale where an observable dominated by strong dynamics is used to set the scale (for instance the masses of hadrons, or the quark-antiquark potential).


\section{Analysis and Results}
\label{sec:Analysis}
\setcounter{equation}{0}

\subsection{$K\to\mu\bar\nu_\mu$, $\pi\to\mu\bar\nu_\mu$ and $B\to\tau\bar\nu_\tau$}
\label{sec:Pell2}

We start with 
the leptonic decay rates $\Gamma(P^+\to\ell^+ \nu_\ell)$, with $P = \{\pi,K,B\}$, and $\ell=\{\mu,\tau\}$.
For
an exhaustive study of $V_{us}$ from $K\to\mu\bar\nu_\mu$ along the lines of the present article, we refer to~Ref.~\cite{Gonzalez-Alonso:2016etj}.
The decay rate for the process $P^-\to\ell^-\bar\nu_\ell$ can be written as
\eq{
\label{eq:Pell2}
\Gamma(P^-\to\ell^-\bar\nu_\ell) = |V_{uq}|^2  \frac{f_P^2\, m_P m_{\ell}^2}{16\pi \tilde v^4}\left ( 1 - {m_\ell^2 \over m_P^2}  \right )^2
\left (1 + \delta_{P \ell} \right)
\left(1+\Delta_{P\ell2}\right)\,,
}
where $f_P$ is the decay constant defined by $\av{0|\bar q\gamma^\mu\gamma_5 u |P^+(k)}= i k^\mu f_P$,
and the quantity $\delta_{P \ell}$ accounts for all electromagnetic corrections in the SM (see e.g.~\cite{Rosner:2015wva}),
\footnote{Factoring out the SM corrections 
induces (tiny) NP $\times$ QED corrections, which may not be the correct ones but which are beyond the current theoretical precision.} 
as well as isospin-breaking corrections if they are not already included in the decay constants~\cite{Giusti:2017dwk}.  
The full and linearized expressions for the NP corrections are given by
\eqa{
\label{eq:delta_Pell2}
\Delta_{P\ell2} &=& \frac{\tilde v^4}{v^4}\, \bigg| 1 +  \epsilon_A^{\ell uq} - {m_P^2 \over (m_u + m_q) m_\ell }\,\epsilon_P^{\ell uq}\bigg|^2 - 1 \nnl[2mm]
& = & 2\ {\rm Re} (\epsilon_A^{\ell uq}) - {2\ m_P^2 \over (m_u + m_q) m_\ell }\ {\rm Re} (\epsilon_P^{\ell uq}) + 4\ \frac{\delta v}{v} + \cO(\Lambda^{-4})\,.
}
The definitions of $\tilde v$ and $\delta v/v$ can be found in~\eref{vtilde}. 
The parameters $\epsilon_A^{\ell uq}$ are $\cO(\Lambda^{-2})$ in the SMEFT expansion, and they are connected to the LEFT  Wilson coefficients at the hadronic scale by:
\eqa{
\epsilon_A^{\ell uq} &\equiv&
-1 - \frac{v^2}{2 V_{uq}}
\Big( \big[L_{\nu edu}^{V,LL} (\mu_q)\big]^*_{\ell\ell q u}
-\big[L_{\nu edu}^{V,LR} (\mu_q)\big]^*_{\ell\ell q u} \Big)
\ ,
\nonumber\\[2mm]
\epsilon_P^{\ell uq} &\equiv& 
- \frac{v^2}{2 V_{uq}}
\Big( \big[L_{\nu edu}^{S,RR} (\mu_q)\big]^*_{\ell\ell q u}
-\big[L_{\nu edu}^{S,RL} (\mu_q)\big]^*_{\ell\ell q u} \Big) \ ,
\label{eq:epsilondef}
}
where $\mu_q=\{2,2,4.3\}\,\rm{GeV}$ for $q=\{d,s,b\}$. 
Using~\eref{semilepWC_LEFT}  we can express $\epsilon_A^{\ell uq}$ and $\epsilon_P^{\ell uq}$  in terms of the LEFT Wilson coefficients at the EW scale, which in turn are matched to the SMEFT Wilson coefficients by means of \eref{semilepWC_SMEFT}. 

Using the input values collected in~\Tab{tabInputs}, 
from the input observable $\Gamma(B \to \tau \nu)$ we obtain
\eq{
\label{eq:Vub}  
|\widetilde V_{ub}|^2 
\equiv  |V_{ub}|^2(1+\Delta_{B\tau2})
= 0.00425 \pm 0.00049\ ,
}
where we neglected the electromagnetic correction (i.e., $\delta_{B \ell}\simeq 0$)
since it induces an effect much smaller that the current experimental sensitivity (see~\Ref{Becirevic:2009aq} for further detail on this issue).
The error in $|\widetilde V_{ub}|$ (12\%) is dominated by the experimental uncertainty on ${\cal B}(B \to\tau \nu_\tau)$.
The value used for the decay constant $f_{B^\pm}$~\cite{Dowdall:2013tga} and quoted in \Tab{tabInputs}  
does not rely on the experimental value for $f_\pi$, c.f. the discussion in \Sec{sec:CKM}.

In a similar fashion we could determine $|\widetilde V_{us}|$ from the observable $\Gamma(K \to \mu \nu)$, which would translate directly into a determination of  the tilde Wolfenstein parameter~$\tilde \lambda$. However, as discussed in Sec.~\ref{sec:CKM}, this choice would lead to a relatively large uncertainty on $\tilde \lambda$, 
which come in particular come from the lattice input for $f_{K}$:
the MILC09 calculation~\cite{Bazavov:2009bb} is
the most precise lattice determination of $f_{K}$ not relying on the pion leptonic width to set the lattice scale, with fairly large uncertainties. 
A more precise value of $\tilde \lambda$ can be obtained by considering instead the ratio $\Gamma(K\to \mu\bar\nu_\mu)/\Gamma(\pi\to \mu\bar\nu_\mu)$,
which allows one to extract the ratio $|\widetilde V_{us}/\widetilde V_{ud}|$ given the ratio of decay constants $f_{K}/f_{\pi}$. 
The latter can be consistently taken from the FLAG average~\cite{Aoki:2019cca}, which combines several lattice determinations for this ratio of decay constants without introducing any uncontrollable dependence on NP via the pion leptonic width~\cite{Aoki:2019cca,Dowdall:2013rya,Carrasco:2014poa,Bazavov:2017lyh}.

In this case we have:
\eq{
\label{eq:K/pi}
\frac{\Gamma(K^-\to\mu^-\bar\nu_\mu)}{\Gamma(\pi^-\to \mu^-\bar\nu_\mu)} =
\frac{|\widetilde V_{us}|^2}{|\widetilde V_{ud}|^2} \frac{f_{K^\pm}^2}{f_{\pi^\pm}^2}
\frac{m_{K^\pm} (1-m_\mu^2/m_{K^\pm}^2)^2}{m_{\pi^\pm} (1-m_\mu^2/m_{\pi^\pm}^2)^2}  (1 + \delta_{K/\pi}) \,,
}
with
\eq{
\frac{|\widetilde V_{us}|^2}{|\widetilde V_{ud}|^2} \equiv \frac{|V_{us}|^2}{|V_{ud}|^2} \, (1+ \Delta_{K/\pi})\,,
}
and
\eqa{
\label{eq:CKM_deltaK/pi}
\Delta_{K/\pi} &=& \frac{1 + \Delta_{K\mu 2}}{1 + \Delta_{\pi\mu 2}} - 1 \nonumber\\
&=& 2\ {\rm Re} ( \epsilon_A^{\mu us} - \epsilon_A^{\mu ud} ) - \frac{2}{m_\mu} 
\left( {\ m_{K^\pm}^2 \ {\rm Re} (\epsilon_P^{\mu us}) \over (m_u + m_s)} - {\ m_{\pi^\pm}^2 \ {\rm Re} (\epsilon_P^{\mu ud}) \over (m_u + m_d) }\right) + \cO(\Lambda^{-4})\,.\qquad
}
Given the inputs in~\Tab{tabInputs}, we find
\eq{\label{eq:VusoVud}
|\widetilde V_{us}/\widetilde V_{ud}| = 0.23131 \pm 0.00050 \,,
}
with a relative error of $0.2\%$, dominated by the uncertainty on the lattice determination on the decay constant ratio.

\subsection{$\Delta M_d$ and $\Delta M_s$}

The mass differences $\Delta M_q$ of neutral $B_q$ mesons ($q=\{d,s\}$) are given by~\cite{Gabbiani:1996hi,Bagger:1997gg}   
\bea
\Delta M_{q} 
&=& |\widetilde V_{tb}\widetilde V_{tq}|^2 \, \frac{m_{B_q}  f_{B_q}^2 m_W^2}{12\pi^2 \tilde v^4} \, B_1^{q}\, S_1(m_b) \,,
\eea
where $B_i^q$ are the so-called \emph{bag parameters}, defined e.g. in Ref.~\cite{Carrasco:2013zta} 
\footnote{The bag parameters $B_i^q$ are the matrix elements $\av{B_q|O_i(\mu)|\bar B_q}$ up to a normalization factor.
With our conventions, $B_i^q$ denote the bag parameters in the renormalisation scheme of~\Ref{Buras:2000if} at a scale $\mu_b=4.3\GeV$, as given Table~2 of~\Ref{Carrasco:2013zta},
and in agreement with the RGE factors used in~\App{app:LEFTrunning}. It is worth mentioning that $f_\pi$ has been used to set the scale in~\Ref{Carrasco:2013zta}; however, this choice induces 
a subleading effect on the determination of the (dimensionless) bag parameters, such that we can safely use this computation given the current uncertainties.}.
The quantities $\widetilde V_{tb}$ and $\widetilde V_{tq}$ have been defined such that:
\beq
\label{eq:CKM_tVtbVtqdef}
|\widetilde V_{tb}\widetilde V_{tq}|^2 \equiv  |V_{tb} V_{tq}|^2 \, (1+\Delta_{\Delta M_{q}})\, , 
\eeq
with the BSM effects contained in
\eqa{
\label{eq:CKM_deltams}
&&
\hspace{-10mm}
\Delta_{\Delta M_{q}} = \frac{\tilde v^4}{v^4}\,\Bigg|
\frac{C_1^{(q)} + \widetilde C^{(q)}_1}{C_{1, \rm SM}^{(q)}} 
+ R_{B_q} \sum_{i=2}^5 \frac{a_i B_i^q}{B_1^{q}}
\frac{C^{(q)}_i}{C_{1, \rm SM}^{(q)}}
+ R_{B_q}\sum_{i=2,3}\frac{a_i B_i^q}{B_1^{q}}
\frac{\widetilde C^{(q)}_i}{C_{1,\rm SM}^{(q)}} 
\Bigg| -1 \nonumber\\
&&=
4 \frac{\delta v}{v} +
{\rm Re} \Bigg[ \frac{C_{1, \rm NP}^{(q)} + \widetilde C^{(q)}_1}{C_{1, \rm SM}^{(q)}}
+ R_{B_q} \sum_{i=2}^5 \frac{a_i B_i^q}{B_1^{q}}
\frac{C^{(q)}_i}{C_{1, \rm SM}^{(q)}}
 + R_{B_q}\sum_{i=2,3}\frac{a_i B_i^q}{B_1^{q}}
\frac{\widetilde C^{(q)}_i}{C_{1,\rm SM}^{(q)}} \Bigg]
+  \cO(\Lambda^{-4}) \ ,\qquad
}
where we have defined $C_{1, \rm NP}^{(q)} = C_1^{(q)} - C_{1, \rm SM}^{(q)}$, and the definition of $\delta v/v$ can be found in~\eref{vtilde}.
Here $a_i = (1,-5/8,1/8,3/4,1/4)$,
and $R_{B_q}\equiv [m_{B_q}/(m_b+m_q)]^2$, where the quark masses $m_b$ and $m_q$ are running $\overline {\rm MS}$ masses at the scale $\mu_b=4.3\GeV$, as are all other scale-dependent parameters in \eref{CKM_deltams}. Numerical values for the bag parameters $B_i^q$ can be found in Table~2 of~\Ref{Carrasco:2013zta}. 
\footnote{We cannot use the more recent calculation of the bag parameters by the MILC collaboration~\cite{Bazavov:2016nty} because they calculate the dimensionful combinations $f_{B_s} \sqrt{\hat{B}_i^q}$, setting the QCD scale with $f_\pi$. The presence of $f_{B_s}$ in their results induces non-negligible effects coming from the use of $f_\pi$, contrary to what happens for the dimensionless quantity computed by ETMC~\cite{Carrasco:2013zta}. This is also the reason why we do not employ more recent computations of $f_{B_s}$, but we rely on Ref.~\cite{Dowdall:2013tga}.}
It is well known that the combination
\eq{
\xi^2 \equiv \frac{f_{B_s}^2 B_1^s}{f_{B_d}^2 B_1^d}
}
is more precisely determined in the lattice than numerator and denominator separately due to the presence of parametric correlations. Often this is exploited by trading one mass difference by the observable $\Delta M_s/\Delta M_d$. Instead, we will take into account these parametric correlations by writing
$f_{B_d}^2 B_1^d = f_{B_s}^2 B_1^s/\xi^2$ in the expression for $\Delta M_d$.

\begin{table}
\centering
\setlength{\tabcolsep}{22pt}
\renewcommand{\arraystretch}{1.5}
\begin{tabular}{@{}ll@{}}
\toprule
$\Gamma(K^+ \to \mu^+ \nu_\mu)  = 3.3793(79) \cdot 10^{-8}~\eV$~\cite{pdg18} &
$\delta_{K/\pi} = -0.0069(17)$~\cite{Cirigliano:2011tm} \\
$\Gamma(\pi^+ \to \mu^+ \nu_\mu)  = 2.5281(5) \cdot 10^{-8}~\eV$~\cite{pdg18} & 
$f_{K^\pm}/f_{\pi^\pm} = 1.1932(19)$ \cite{Aoki:2019cca,Dowdall:2013rya,Carrasco:2014poa,Bazavov:2017lyh} \\
$\Gamma (B^+ \to \tau^+ \nu_\tau)  = 4.38 (96) \cdot 10^{-8}~\eV$~\cite{pdg18}  &
$f_{B^\pm} = 184(4)~\MeV $ \cite{Dowdall:2013tga}\\
$\Delta M_d  = 3.333 (13) \cdot 10^{-10}~\MeV$~\cite{pdg18}  &  
$f_{B_s} = 224(4) \MeV$ \cite{Dowdall:2013tga} \\
$\Delta M_s  = 1.1688 (14) \cdot 10^{-8}~\MeV $~\cite{pdg18}   
& $B_1^s = 0.86(3)$ \cite{Carrasco:2013zta} \\
$S_1(m_b) \simeq 1.9848$ \ ({\it cf.}~App.~\ref{app:LEFTSMEFTmatching})
&$\xi = 1.206(17)$ \cite{Aoki:2019cca,Aoki:2014nga,Bazavov:2016nty} \\
$\tilde v = 246.21965(6)$~GeV~\cite{pdg18} & 
$m_W = 80.379(12)~\GeV$ \cite{pdg18}  \\
$m_\mu = 105.6583745(24)~\MeV$~\cite{pdg18}     &   
$m_\tau = 1.77686(12)~\GeV$  \cite{pdg18}  \\
$m_{\pi^\pm} = 139.57061(24)~\MeV$~\cite{pdg18} & 
$m_{K^\pm} = 493.677(16)~\MeV$~\cite{pdg18}  \\
$m_{B^\pm} = 5.27932(14)~\GeV$ \cite{pdg18}  &  
$m_{B_d} = 5.27963(15)~\GeV$ \cite{pdg18} \\
$m_{B_s} = 5.36689(19)~\GeV$ \cite{pdg18} & 
 \\
\bottomrule
\end{tabular}
\caption{Set of inputs used in the numerical analysis.
}
\label{tabInputs}
\end{table}

The contributions of various effective operators are parametrized by $C_i^{(q)}, \widetilde C^{(q)}_i$. In the SM only $C_{1}^{(q)}$ is generated, with $C_{1,\rm SM}^{(q)}$ known at NLO in QCD, and given in \eref{C1_SM}.
The relation of the coefficients $C_i^{(q)}, \widetilde C^{(q)}_i$ to the LEFT Wilson coefficients at $\mu_{\rm EW}$ is defined in \eref{neutralWC_LEFT}, which in turn can be matched to the SMEFT Wilson coefficients by means of \eref{neutralWC_SMEFT}.
For the NP contributions, we give explicitly in~\eref{neutralWC_SMEFT} only the tree-level matching conditions between the LEFT and the SMEFT, but higher order corrections can be included trivially, once they are
known (e.g.~\cite{Bobeth:2017xry}). This might be relevant since non-log-enhanced GIM-violating contributions may be numerically large in observables such as~$\Delta M_q$. This is an important issue to keep in mind.

Using the numerical inputs from~\Tab{tabInputs}, we obtain
\beq
\label{eq:VtbVtq}
|\widetilde V_{tb} \widetilde V_{td}|  = 0.00851 \pm 0.00025 \,,
\qquad \text{and} \qquad
|\widetilde V_{tb} \widetilde V_{ts}|  = 0.0414 \pm 0.0010 \,.
\eeq
The errors in $|\widetilde V_{tb} \widetilde V_{td}|$ (2.9\%) 
and $|\widetilde V_{tb} \widetilde V_{ts}|$ (2.5\%) 
are both dominated by the uncertainties of the $f^2_{B_q} B_1^q$ combinations. 
Again, the $f_{B_s}$ value~\cite{Dowdall:2013tga} 
quoted in \Tab{tabInputs} does not rely on 
the experimental value for $f_\pi$, according to the discussion in \Sec{sec:CKM}.

\subsection{Summary and Results}

To summarise, we have obtained the following numerical constraints on the tilde CKM elements:
\begin{align}
|\widetilde V_{us}/\widetilde V_{ud}| &= 0.23131 \pm 0.00050\,,&
|\widetilde V_{ub}| & =  0.00425 \pm 0.00049\,,  
\nonumber\\
|\widetilde V_{tb} \widetilde V_{td}| & =  0.00851 \pm 0.00025\,,&
|\widetilde V_{tb} \widetilde V_{ts}| & =  0.0414 \pm 0.0010\,, 
\label{eq:CKM_tvlimits}
\end{align}
with negligible correlations, except for the last two elements that present a correlation of $+87\%$. Moreover, we have identified the new physics contributions to the above quantities, which can be found in Eqs.~\eqref{eq:delta_Pell2},~\eqref{eq:CKM_deltaK/pi}, and~\eqref{eq:CKM_deltams}.

It is now possible to write our results in terms of the tilde Wolfenstein parameters $\widetilde W_i$, making use of the following relations:
\bea
|\widetilde V_{us}/\widetilde V_{ud}| &=& \tilde \lambda + \frac12 \tilde\lambda^3 + \frac38 \tilde\lambda^5 + \cO(\lambda^7)\,,\nnl
|\widetilde V_{ub}| &=&  \tilde A \sqrt{\tilde\rho^2 + \tilde \eta^2} \, \Big[ \tilde \lambda^3  + {1 \over 2} \tilde \lambda^5  + \cO(\lambda^7) \Big] \nnl
|\widetilde V_{tb} \widetilde V_{td}| &=& \tilde \lambda^3 \tilde A \sqrt{(1 - \tilde \rho)^2 + \tilde \eta^2} + \cO(\lambda^7)\,,\nnl
|\widetilde V_{tb} \widetilde V_{ts}| &=&  \tilde \lambda^2  \tilde A  - \frac12 \tilde \lambda^4  \tilde A  (1 - 2 \tilde \rho) + \cO(\lambda^6)\,.
\eea
Given these definitions, one can translate \eref{CKM_tvlimits} into correlated constraints on the CKM parameters $\widetilde W_i$. 
We find the following results:
\eq{
\label{eq:tildeW}
\bvec 
\tilde \lambda = \lambda + \delta \lambda  \\ 
\tilde A = A + \delta A  \\ 
\tilde \rho = \barrho +  \delta \barrho   \\ 
\tilde \eta = \bareta +  \delta \bareta  
\evec
= 
\bvec
0.22537 \pm 0.00046 \\
0.828 \pm 0.021 \\
0.194 \pm 0.024 \\
0.391 \pm 0.048 \evec, 
\quad 
\rho = \left ( \ba{rrrr}
1\  & -0.16 & \ 0.05 & -0.03 \\ 
\cdot & 1\  & -0.25 & -0.24 \\ 
\cdot & \cdot & 1\  & 0.83 \\ 
\cdot & \cdot &  \cdot  & 1\   
\ea \right )\,.
}
Our choice of input observables leads to moderate correlations between the numerical values of the tilde parameters, except in the $(\trho,\teta)$ case. 
Note also that the accuracy of the determination of $\tilde \lambda$ justifies retaining $\cO(\lambda^5)$ terms in \eref{CKM_vckm}, and neglecting $\cO(\lambda^6)$ ones.

At leading order in the EFT expansion the NP shifts to the Wolfenstein parameters $\delta W_j$ 
correspond to the following combinations of NP Wilson coefficients:
\beq
\label{eq:deltaWmatricial}
\bvec 
\delta \lambda \\ 
\delta A  \\ 
\delta \barrho   \\ 
\delta \bareta 
\evec 
= 
M(\tlambda,\tA,\trho,\teta) 
\bvec 
\Delta_{K/\pi}  \\ 
\Delta_{B\tau2} \\ 
\Delta_{\Delta M_{d}}  \\ 
\Delta_{\Delta M_{s}}
\evec~,
\eeq
where 
$\Delta_{K/\pi}, \Delta_{B\tau2}, \Delta_{\Delta M_{d}}$, and $\Delta_{\Delta M_{s}}$ are the (linearized) NP contributions to the four chosen observables, which can be found 
in Eqs.~\eqref{eq:delta_Pell2},~\eqref{eq:CKM_deltaK/pi}, and~\eqref{eq:CKM_deltams}, and
the matrix $M$ is given by
\beq
\label{eq:deltaWmatrix}
M = \left ( 
\ba{cccc}
\frac12 \tlambda - \frac12 \tlambda^3 & 0 & 0 & 0 \\ 
-\tA + \tA \tlambda^2 + c\,\tA\, \tlambda^4& 
-c\, e \,\tA& 
b\, e \,\tA& 
\frac12 \tA - a\, e \,\tA\\
a - b\tlambda^2 + \frac{c\,(5 - 4\trho)}{2}\tlambda^4  & 
c( 1 - 2 a\, e ) & 
-b( 1 - 2 a\, e ) & 
a( 1 - 2 a\, e )  \\
-\frac{d}{2\teta} + \frac{b\,\trho}{\teta}\tlambda^2 - \frac{c\,(2d+3(\trho - 1))}{2\teta}\tlambda^4 & 
\frac{c}{\teta}( 1 - \trho + d\, e ) & 
\frac{b}{\teta}( \trho - d\, e ) & 
-\frac{d}{2\teta}( 1 - 2 a\, e )   
\ea \right )
+ \mathcal{O}(\tlambda^6)\,,
\eeq
with
\beq
a \equiv \frac{1-2\trho}{2}\ ,\quad
b \equiv \frac{\teta^2 + (1-\trho)^2}{2}\ ,\quad
c \equiv \frac{\teta^2 + \trho^2}{2}\ ,\quad
d \equiv \teta^2-\trho^2+\trho\ ,\quad
e \equiv \tlambda^2(1-a\tlambda^2)\,.
\eeq 
The numerical value of $M$ is given by
\beq
\label{eq:Mnumerical}
M(\tilde\lambda,\tilde A,\tilde\rho,\tilde\eta) = \left (
\ba{rrrr}
0.1070(2) & 0\qquad & 0\qquad & 0\qquad \\ 
-0.786(20) & -0.0040(9) & 0.0167(6) & 0.402(10) \\ 
0.286(24) & 0.094(22) & -0.390(10) & 0.296(23) \\ 
-0.385(18) & 0.200(19) & 0.184(10) & -0.384(19)
\ea \right )~.
\eeq
Eqs.~\eqref{eq:tildeW}-\eqref{eq:Mnumerical} represent the main results of this work.

\begin{table}
\centering
\setlength{\tabcolsep}{22pt}
\renewcommand{\arraystretch}{1.5}
\begin{tabular}{@{}lll@{}}
\toprule
CKMfitter (SM) \cite{Charles:2004jd} & UTfit (SM) \cite{Bona:2007vi} & This work (SMEFT) \\
\midrule
$\lambda  = 0.224747^{+0.000254}_{-0.000059}$   &  
$\lambda  = 0.2250 \pm 0.0005$ &  
$\tilde \lambda = 0.22537 \pm 0.00046$  \\
$A = 0.8403^{+0.0056}_{-0.0201}$    &  
$A = 0.826 \pm 0.012$    &  
$\tilde A = 0.828 \pm 0.021$    \\
$\bar \rho = 0.1577^{+0.0096}_{-0.0074}$  &  
$\barrho = 0.148 \pm 0.013$   &  
$\tilde \rho = 0.194 \pm 0.024$  \\
$\bar \eta = 0.3493^{+0.0095}_{-0.0071}$    &  
$\bareta = 0.348 \pm 0.010$   &  
$\tilde \eta = 0.391 \pm 0.048$  \\
\bottomrule
\end{tabular}
\caption{Results for the Wolfenstein parameters $\widetilde W_i$ extracted here compared to the Wolfenstein parameters
extracted from the canonical SM fits.}
\label{tabResults}
\end{table}

\Tab{tabResults} summarises our results for the Wolfenstein parameters in the presence of NP, and compares them to the results of the canonical SM fits.
As could be expected, our procedure of using only four input  observables to determine the four Wolfenstein parameter leads to some loss of accuracy in the limit where BSM corrections are absent, compared to the SM fits using a much larger set of observables.\footnote{The difference of precision between our determination of $\tilde\lambda$ and the SM CKMfitter determination of $\lambda$ is partially due to the use of a larger set of constraints in the latter case, but mainly due to current internal tensions between some of the constraints on $\lambda$ (within the SM), which generate a smaller error on $\lambda$ in the CKMfitter statistical approach. Compared to UTfit, we obtain a more precise $\tilde{\lambda}$ because we use the new FLAG average for $f_K/f_\pi$~\cite{Aoki:2019cca}.} 
Nevertheless, in most physical applications the error bars of our tilde parameters will be anyway a subleading effect compared to other sources of experimental or theoretical uncertainties (as illustrated in a few concrete examples in the next section).
On the other hand our tilde parameters can be consistently used for generic NP frameworks described by the SMEFT, unlike the results of the SM fits. 
We remark that our input observables allow other solutions than the one displayed in \eref{tildeW}; in particular, there is another solution with the opposite sign of $\tilde \eta$. 
This discrete ambiguity will lead to ``mirror solutions" in global fits,  where the CKM parameters differ significantly from the ones obtained in the  SM context, but the resulting shift of precisely measured flavour observables is canceled by a relatively large (and fine-tuned) contribution from  SMEFT Wilson coefficients.
In this article we will not discuss the mirror solutions any further, 
and focus on the SM-like solution in \eref{tildeW} where the NP effects are subleading compared to the SM contributions.  

With the likelihood function in~\eref{tildeW} we find the following $1\sigma$ intervals for the elements of the tilde CKM matrix defined in \eref{tildeVjk}:
\eq{
{
\footnotesize
\widetilde V = \left (
\ba{ccc}
0.97428(11) & 0.22537(46) & 0.00189(23) - i~0.00380(45)\\ 
-0.22524(46) - i~0.000156(19) & 0.97340(12) & 0.0421(11) \\
0.00764(34) -i~0.00370(44) &  -0.0414(10) -i~0.00083(10)&  0.999114(45)
\ea
\right ).
}
\label{eq:tildeVjknum}
}
We do not give here the (non-trivial) correlations between the various tilde CKM elements, but they are encoded in the likelihood in~\eref{tildeW}.
The NP effects absorbed in these CKM elements should be taken into account through the method described in~\sref{CKM}; see \sref{hadronicw} for an example.
The numerical form of the matrix $\widetilde V$ is given here for illustration
purposes only: ideally one should always express all CKM input in terms of Wolfenstein parameters if one wishes to use the approach and results of this paper appropriately, including correlations.


\section{Applications}
\label{sec:applications}
\setcounter{equation}{0}

In this section we discuss through a few examples how to use the tilde parameters to analyze, {\it in a consistent fashion},  other flavour processes and set bounds on NP. 

\subsection{Leptonic decays of pions and $D$ mesons}

Consider the pion decay $\pi \to \mu \nu$. 
The goal is to compare the precisely measured branching fraction to the SM predictions so as to place constraints on effective interactions beyond the SM. 
The SM prediction is proportional to $|V_{ud}|$.
In the presence of generic NP we cannot use $|V_{ud}|$ determined by fits performed in the SM context, such as the ones provided by CKMfitter or the PDG, as the observables used in those analyses may themselves be affected by new physics, which might even have the same underlying effective operators.
Instead, we can use our results in \eref{tildeW} where the NP effects have been absorbed into the definition of the tilde parameters. 
All we need to do is to express the theoretical prediction for ${\cal B}(\pi \to \mu \nu)$ in terms of $\tilde \lambda$ defined in \eref{deltaWmatricial}.

The $\pi \to \mu \nu$ decay width can thus be written as
\beq
\Gamma(\pi \to \mu \nu)  = 
\left | 1  - {\tilde \lambda^2 \over 2}   - {\tilde \lambda^4 \over 8}   \right |^2 
{f_{\pi^\pm}^2 m_{\pi^\pm} m_\mu^2  \over 16 \pi \tilde v^4} \left(1 - \frac{m_\mu^2}{m_{\pi^\pm}^2} \right)^2
\left ( 1 + \delta_{\pi \mu} \right )
\left [ 1  +  \widetilde \Delta_{\pi\mu 2} \right ] \,,  
\eeq
where the decay constant is $f_{\pi^\pm} = 130.2(8) \MeV$ (from the average of Ref.~\cite{Aoki:2019cca} with a lattice scale set using QCD observables~\cite{Follana:2007uv,Bazavov:2010hj,Arthur:2012yc}), the electromagnetic radiative corrections
are encoded in
$\delta_{\pi \mu} = 0.0176(21)$~\cite{Rosner:2015wva},
and $\widetilde\Delta_{\pi\mu 2}$ is given by
\eq{
\label{eq:APP_deltapiellnu}
\widetilde \Delta_{\pi\mu 2} =  2\,{\rm Re} (\epsilon_A^{\mu ud}) 
-\frac{2 m_{\pi^\pm}^2 }{(m_u + m_d)m_\mu}{\rm Re}(\epsilon_P^{\mu ud})
+4 \frac{\delta v}v 
+2 \tilde \lambda (1 +\tilde \lambda^2) \delta \lambda  + \cO(\Lambda^{-4},\tilde \lambda^6)\ . 
}
The NP quantities $\epsilon_{A,P}^{\mu ud}$, $\delta v/v$ and $\delta\lambda$ have been defined in Eqs.~(\ref{eq:epsilondef}), (\ref{eq:vtilde}), (\ref{eq:deltaWmatricial}). 
The terms proportional to $\delta\lambda$ are due to NP affecting the observables  used to determine the CKM parameters in our approach;
note that they depend on the same Wilson coefficients that also enter into $\epsilon_{A,P}^{\mu ud}$.
Their effect is to change the linear combination of Wilson coefficients $\tilde \Delta_{\pi\mu 2}$ probed by $\pi \to \mu \nu$ decays.     
Given the current experimental measurement,  
$\B(\pi \to \mu \nu) = 0.9998770(4)$~\cite{pdg18}, 
combined with $\tau_\pi = 2.6033(5)\cdot 10^{-8} s$, 
we obtain the following constraints on the linear combinations of Wilson coefficients in~\eref{APP_deltapiellnu}:   
\beq
\label{eq:piondecay}
\widetilde \Delta_{\pi\mu 2}  = 0.004 \pm 0.013. 
\eeq 
The error is totally dominated by the lattice uncertainty on $f_{\pi^\pm}$.
The error of our determination of $\tilde \lambda$ in \eref{tildeW} is completely negligible for this constraint.   

Up to small $\cO(\lambda^4)$ corrections, the CKM elements $V_{ud,us,cd,cs}$ are only functions of $\lambda$ in the Wolfenstein parameterization. Thus, besides pion decays, there is a long list of observables which are only functions of $\lambda$ and NP parameters. 
A global fit to $d\to u\ell\nu$ and $d\to u\ell\nu$ transitions
was performed  Ref.~\cite{Gonzalez-Alonso:2016etj}, where simultaneous constraints were derived on $\lambda$ and the relevant LEFT Wilson coefficients. We note that such a global approach obviates the need to define the tilde CKM parameters;\footnote{More precisely, the definition of $\tilde{\lambda}$ is implicit in Ref.~\cite{Gonzalez-Alonso:2016etj}. Indeed, $\tilde{V}_{ud}$ and $\tilde{V}_{us}$ are introduced (although the definition is different to this work) and the use of CKM unitarity amounts to defining the corresponding $\tilde{\lambda}$.} however, so far it was realised only for observables depending on $\lambda$, and extending it to the full set of the CKM parameters will involve considerable technical difficulties. The approach in this article bypasses this problem when setting constraints from individual observables, as has been exemplified here.

Going beyond the analysis of Ref.~\cite{Gonzalez-Alonso:2016etj}, we consider the decay $D \to \ell \nu$.
Analogously to  $\pi \to \ell \nu$, we write
\beq
\Gamma(D \to \ell \nu)  = 
| \tilde \lambda |^2 
{f_{D^\pm}^2 m_{D^\pm} m_\ell^2  \over 16 \pi \tilde v^4} \left(1 - \frac{m_\ell^2}{m_{D^\pm}^2} \right)^2
\left ( 1 + \delta_{D \ell} \right )
\left [ 1  + \widetilde \Delta_{D\ell 2} \right ] \,,  
\eeq 
where $f_{D^\pm} = 212.7\,(6) \MeV$~\cite{Bazavov:2017lyh},
and $\widetilde \Delta_{D\ell 2}$ is given by 
\beq
\label{eq:APP_deltaDellnu}
\widetilde \Delta_{D\ell 2} =  
2 \, {\rm Re} (\epsilon_A^{\ell cd}) - {2\,m_{D^\pm}^2 \over (m_c + m_d)m_\ell } \ {\rm Re} (\epsilon_P^{\ell cd}) + 4 {\delta v \over v}
- 2 \frac{\delta \lambda}{\tilde \lambda}  + \cO(\Lambda^{-4},\tilde \lambda^4). 
\eeq
where the NP quantities $\epsilon_{A,P}^{\ell cd}$ can be found again in \eref{epsilondef}.
The experimental measurement
$\B (D \to \mu \nu) = 3.74\,(17) \cdot 10^{-4}$~\cite{pdg18} combined with the lifetime $\tau_{D^\pm} = 1.040(7) \cdot 10^{-12} s$, and $\delta_{D \mu} = 0.007\,(6)$~\cite{Bazavov:2017lyh},
results in the following constraint: 
\beq
\label{eq:APP_deltaDmunuNum}
\widetilde \Delta_{D\mu 2}  = -0.089  \pm 0.043, 
\eeq
showing a small $2.1\,\sigma$ tension with the SM.
Our analysis affects the NP interpretation of this result, adding the $\delta \lambda$ term in \eref{APP_deltaDellnu}.
In this case this correction is enhanced by $\tilde \lambda$ (unlike for $\pi \to \ell \nu$ where it is suppressed). 

One comment about the lattice input for $f_{D^\pm}$ is in order in connection with the discussion in \sref{CKM}.  
The value of $f_{D^\pm}$ obtained in~\cite{Bazavov:2017lyh} is normalised to the PDG value $f_\pi^{\rm exp}$, which in principle propagates the NP contribution in $\pi \to \mu \nu$ into the $D \to \mu \nu$ constraints.
However, the comparison of $f_\pi^{\rm lattice}$ and $f_\pi^{\rm exp}$ limits these NP effects to the $1\%$ level, {\it cf.}~\eref{piondecay}, 
which makes them subdominant compared to the  $~4\%$ error in \eref{APP_deltaDmunuNum} mainly due to
the experimental uncertainty on~\mbox{$\B(D \to \mu \nu)$}.
To avoid this issue altogether, lattice collaborations should quote $f_{D^\pm}$ 
setting the scale using a QCD-dominated observable free of NP. 
These considerations may be relevant in the future, when the experimental error on $\B(D \to \mu \nu)$ is reduced.

\subsection{Exclusive hadronic $W$ decays}
\label{sec:hadronicw}

We now consider the processes $W \to u_j d_k$, where $u_j = (u,c)$ and $d_k = (d,s,b)$ denote particular quark flavours. 
We assume that flavour tagging allows one to separate the distinct exclusive hadronic $W$~decays, and that it is possible to measure the partial widths  with a good precision. 
In the SM, the measurement of $\Gamma(W \to u_j d_k)$ can be interpreted as an alternative probe of the CKM element~$V_{jk}$. 
Beyond the SM, 
the coupling strength of the $W$ boson to quarks may be affected by new physics. 
In the SMEFT, the leading order effects can be described by the vertex corrections $\delta g^{Wq}_L$ to the couplings between the left-handed quarks and  $W$:
\beq 
\label{eq:APP_Lwqq}
\cL_{\rm SMEFT} \supset 
 {\tilde g_L \over \sqrt 2}W^{\mu+} \bar u_{Lj} \gamma_\mu 
 \left(V_{jk} +  \big[\delta g^{W q}_L \big]_{jk}  \right) d_{Lk} +\hc
\eeq 
The right-handed vertex correction $\delta g^{Wq}_R$ does not affect $W$ decays at $\cO(\Lambda^{-2})$, and will be neglected in this discussion. 
$\tilde g_L$, $\tilde g_Y$ are the $SU(2)_L \times U(1)_L$ gauge couplings extracted from the EW input observables $\alpha$, $m_Z$ in the presence of dimension-6 operators, in analogy to $\tilde v$ extracted from $G_F$ discussed in \sref{tildev}.
The left-handed vertex correction is related to the parameters in the Warsaw basis as~\cite{Falkowski:2017pss}
\bea
\label{eq:APP_dgwqlwarsaw}
\big[\delta g^{Wq}_L \big]_{jk}  & =&   
[C^{(3)}_{H q}]_{jl} V_{lk}   +  {\tilde g_L^2 \tilde v^2 \over \tilde g_L^2 - \tilde g_Y^2} 
\bigg [  - {\tilde g_Y \over \tilde g_L} C_{HWB} 
-  {1 \over 4} C_{HD}  
\nnl && 
+  {1 \over 4 }[C_{\ell \ell}]_{e\mu\mu e} + {1 \over 4}[C_{\ell \ell}]_{\mu e e \mu}
- {1 \over 2} [C^{(3)}_{H \ell } ]_{ee}  - {1 \over 2 } [C^{(3)}_{H \ell } ]_{\mu \mu} 
\bigg ]V_{jk} + \cO(\Lambda^{-4}) \,. \quad
\eea
Naively, from \eref{APP_Lwqq} one could conclude that each exclusive $W$ decays probes simply~$\big[\delta g^{W q}_L \big]_{jk}$ and thus constrains the particular combination of the Wilson coefficients given in \eref{APP_dgwqlwarsaw}. 
However, to constrain new physics, the experimentally measured partial width has to be compared with the corresponding SM prediction. 
The exclusive decay widths predicted in the SM depend on the numerical value for the specific CKM matrix element $V_{jk}$ extracted from experiment, which in turn may be affected by NP. 
This effect can be disentangled in our scheme by trading $V_{jk}$ in~Eqs.~(\ref{eq:APP_Lwqq}) and~(\ref{eq:APP_dgwqlwarsaw}) for the tilde CKM element $\widetilde V_{jk}$ defined in~\eref{tildeVjk},
i.e. $V_{jk} \to \tilde V_{jk} -\delta V_{jk}$.
We then have
\beq
{\Gamma(W \to u_j d_k) \over \Gamma(W \to u_j d_k)_{\rm SM} } = 
1 +  2 \, \re  \left ( 
{\big[\delta g^{W q}_L\big]_{jk}  -  \delta V_{jk} \over \widetilde V_{jk} } \right ) ,  
\eeq
where $\Gamma(W \to u_j d_k)_{\rm SM}$ is the SM prediction calculated using the numerical values of the tilde CKM elements in~\eref{tildeVjknum}. 
In this approach, the combination probed by the exclusive decay $W \to u_j d_k$ is $\big[\delta g^{W q}_L\big]_{jk} -  \delta V_{jk}$. 
The CKM shifts relevant for $W$ decays, up to $\cO(\Lambda^{-4})$ and at leading order in $\tilde \lambda$, are given by
\eqa{
\label{eq:deltavjk}
\delta V_{ud} &= & \delta V_{cs}  =  - \tlambda \, \delta\lambda
+ \cO{(\tlambda^4)} \ ,
\nnl[2mm]
\delta V_{us} & = & - \delta V_{cd} = \delta\lambda 
+ \cO{(\tlambda^5)} \ ,
\nnl[2mm]
\delta V_{ub} &=& 3\tA \tlambda^2 (\trho -i\teta)\,\delta\lambda
+ \tlambda^3 (\trho -i\teta)\,\delta A
+ \tA \tlambda^3 (\delta\rho -i\delta\eta)
+ \cO{(\tlambda^5)} \ ,
\nnl[2mm]
\delta V_{cb} &=& 2 \tA\,\tlambda\,\delta\lambda + \tlambda^2\,\delta A
+ \cO{(\tlambda^6)} \ .
}
The corrections to the Wolfenstein parameters $\delta \lambda$, $\delta A$, $\delta \bareta$ and $\delta\barrho$ in terms of EFT Wilson coefficients are defined in  \eref{deltaWmatricial}.
If needed, (more lengthy) $\cO(\tilde \lambda^4)$ corrections to the expressions in \eref{deltavjk} can be easily calculated, but we do not list them here for the sake of simplicity.
 
 The data on exclusive $W$ decays are presently very limited. 
 We are only aware of the Delphi measurement of $\Gamma(W \to c s)$~\cite{Abreu:1998ap} with the relative precision of $\sim 40$\%.  
 However, progress in flavour tagging should enable more precise measurements of $\Gamma(W \to u_j d_k)$ in the near future. 
 For example, Ref.~\cite{Harrison:2018bqi} argues that the measurement of $\Gamma(W \to c b)$ with the relative precision of order $\sim 15$\% should be possible in ATLAS or CMS using the existing data sets. 
In  the scheme proposed in the present article, an LHC measurement of  $\Gamma(W \to c b)$ probes not only the vertex correction  $\big[\delta g^{W q}_L\big]_{cb}$, but also 4-quark operators affecting  the $B_s$  meson mass difference  $\Delta M_s$. 
 Our formalism allows for a consistent interpretation of these measurements as model-independent constraints in the SMEFT.

\subsection{A $Z'$ model for $b\to s\ell\ell$ anomalies}

To close this section, we discuss an
application of our formalism in the context of a toy model addressing  $b\to s\ell\ell~(\ell=e,\mu)$ anomalies, and in particular the  ratios $R_{K}$ and $R_{K^*}$ violating lepton-flavour universality \cite{Hiller:2003js,Aaij:2014ora,Aaij:2017vbb,Bifani:2018zmi}. 
Let us consider a simple BSM toy model featuring a massive $Z'$ boson coupled in an $SU(3)\times SU(2) \times U(1)$ invariant way to left-handed  $b$ and $s$ quarks and to left-handed muons:  
\begin{equation}
{\cal L} \supset g_{bs} Z'_\rho  \left(
\bar q_2 \gamma^\rho q_3  + \hc  \right) 
- g_{\mu \mu} Z'_\rho   \bar \ell_2 \gamma^\rho \ell_2
. 
\label{eq:APP_zpmodel}
\end{equation}
Here $\ell_2 = (\nu_\mu,\mu_L)$ is the second-generation lepton doublet, 
and $q_2 = (V^\dagger_{2x} u_{L,x},s_L)$, $q_3 = (V^\dagger_{3x} u_{L,x},b_L)$ are the second- and third-generation quark doublets in the down-type basis. 
Integrating out  $Z'$ yields new contributions to four-fermion contact interactions in the effective theory below the scale $m_{Z'}$.  
 In particular, we generate a new contribution to the effective interaction  $(\bar{b}_{L} \gamma_\rho s_{L})(\bar \mu_{L} \gamma^\rho \mu_{L})$ that adds to the SM loop-level contribution and may help to explain the $R_{K^{(*)}}$ anomalies, as pointed out by several independent analyses~\cite{Capdevila:2017bsm,Altmannshofer:2017yso,DAmico:2017mtc,Hiller:2017bzc,Geng:2017svp,Ciuchini:2017mik,Celis:2017doq,Alok:2017sui,Hurth:2017hxg}. Our model corresponds to the scenario  $C_{9\mu}^{\rm NP} = - C_{10\mu}^{\rm NP}$ in the formalism  of the effective Hamiltonian used in these references.
In the LEFT notation, we have 
\begin{equation}
    \Delta [L_{ed}^{V,LL}(m_b)]_{\mu \mu s b} 
    = (1+\rho_{ed}^{V,LL})\,{g_{bs} g_{\mu \mu} \over  m_{Z'}^2 }
    = { 1.00 \pm 0.21 \over (31.3~\text{ TeV})^2}~,
\end{equation}    
on top of the SM one-loop contribution $[L_{ed}^{V,LL}(m_b)]_{\mu \mu s b}^{\rm SM}  \approx  (12~\text{ TeV})^{-2}$. We have used the results of Ref.~\cite{Capdevila:2017bsm} for the best fit to $b\to s\ell\ell$ flavour observables. The small correction $\rho_{ed}^{V,LL}$ takes into account the running from the $Z'$ mass to the $b$ mass. 

This constrains one combination of the three toy-model parameters $g_{bs}$, $g_{\mu\mu}$, and $m_{Z'}$. 
In addition, 
other operators generated by integrating out the $Z'$ boson lead to further constraints on these parameters. 
First, the low-energy theory contains new contributions to four-lepton interactions 
that can be probed by the  trident muon production in neutrino scattering~\cite{Geiregat:1990gz,Mishra:1991bv,Altmannshofer:2014pba}. Namely
\bea
\Delta [L_{\nu e}^{V,LL}(\mu_{EW})]_{\mu \mu \mu \mu}
=   - (1+\rho_{\nu e}^{V,LL}){g_{\mu \mu}^2 \over  m_{Z'}^2}
= {-0.02 \pm 0.21 \over (246~\text{GeV})^2}~,
\eea
where $\rho_{\nu e}^{V,LL}$ accounts for the running from $m_{Z'}$ to the EW scale. The numerical value in the RHS was taken from the global fit in Ref.~\cite{Falkowski:2017pss}.

Furthermore, we generate  
the contribution to the $\Delta F=2$ operator responsible for $B_s$ mixing
$C_1^{(s)} - C_{1,\rm SM}^{(s)} = - {g_{bs}^2 / (2 m_{Z'}^2)}$,
which adds up to the loop SM contribution 
$C_{1,\rm SM}^{(s)}$ defined in \eref{C1_SM}.
The $Z'$ contribution to the $B_s$ meson mass difference reads
\beq
\Delta_{\Delta M_s} = \re {C_1^{(s)}  - C_{1,\rm SM}^{(s)}  \over C_{1,\rm SM}^{(s)} }
\approx 2.3 \left ( g_{bs} \over 10^{-3} \right )^2 \left (100~\gev \over m_{Z'} \right )^2 \left (1 + \rho_1^{(s)} \right) \, , 
\eeq 
where $\rho_1^{(s)}$ accounts for RG running from $m_{Z'}$ to the EW scale.
One may be tempted to derive  a constraint on the combination $g_{bs}^2/m_{Z'}^2$ using the measured value of $\Delta M_s$~\cite{pdg18} together with its SM prediction calculated using the CKM elements extracted within the SM~\cite{DiLuzio:2017fdq}, 
namely $\Delta_{\Delta M_s} = \left( \Delta M_s^{\rm{exp}} - \Delta M_s^{\rm{SM}} \right) / \Delta M_s^{\rm{SM}}$.  
There is however one conceptual difficulty. 
The SM prediction for $\Delta M_s$ crucially depends on the numerical value of the CKM matrix element $V_{ts}$. 
In the SM context this value is extracted from a global fit to multiple observables (including $\Delta M_s$) and it may be shifted in the presence of NP contributions. 
Our approach allows us to solve this conundrum.  
Since the $\Delta M_s$ measurement  has been selected as one of our input observables,  it serves as an input to fix the tilde Wolfenstein parameters and  by itself it does not constrain new physics.  
Of course, this does not mean that there is no possibility to probe $C_1^{(s)}$, but we need to use other observables than $\Delta M_s$
such as the $B \to D^{(*)} \ell \nu~(\ell=e,\mu)$ decays.  
Following the analysis of Ref.~\cite{Jung:2018lfu}  in the SM limit, $B \to D^{*} \ell \nu$ yields the 68\%~CL constraint on the  CKM parameter $|V_{cb}| = (3.90 \pm 0.07) \times 10^{-2}$,   
while from  $B \to D \ell \nu$ one obtains  $|V_{cb}| = (3.96 \pm 0.09) \times 10^{-2}$.
These extractions happen to be valid also in our model, since it does not introduce NP contributions to $b \to c\ell\nu$  transitions. Following the scheme proposed in this article, we can then relate the extracted $|V_{cb}|$ to its tilde value, which gives us the following constraint on the model:
\bea
|V_{cb}|  & = &  A \lambda^2 + \cO( \lambda^6)
 = \tilde A \tilde \lambda^2 \left [1 - 2{ \delta \lambda \over \tilde \lambda} - {\delta A \over \tilde A} \right ]  + \cO( \lambda^6)
=   \widetilde V_{cb} \left [ 1 -  0.485\,  \Delta_{\Delta M_s}  \right ]  + \cO( \lambda^6)
 \nnl  & \approx &  
 \widetilde V_{cb} \left [ 1  - 1.1  \left ( g_{bs} \over 10^{-3} \right )^2 \left (100~\gev \over m_{Z'} \right )^2 \right ] + \cO( \lambda^6)~. 
\eea 
where we used \eref{deltaWmatricial} to relate $\delta A$ to $\Delta_{\Delta M_s}$. We used as well that $\Delta_{K/\pi}$ (and thus also $\delta \lambda$), $\Delta_{B\tau2}$, and $\Delta_{\Delta M_{d}}$ are zero in this model. 
Illustrative limits on our model are shown in \fref{bszprimereloaded} in the case $m_{Z'} = 100 \GeV$.
Together with the trident constraints,  $B \to D^{(*)} \ell \nu$  leaves a corner of allowed parameter space in the $g_{\mu\mu}$-$g_{bs}$ plane where the $Z'$ couplings are small enough. 
In our approach, 
$B \to D^{*} \ell \nu$ 
provides the strongest constraint on $g_{bs}$, 
which is however weaker than what would be found if we (incorrectly) used  $\Delta M_{s}$ to set limits.   
The outcome of the two approaches could differ even more significantly in more general situations, for example when an extra gauge boson also generates $bc \ell \nu$ effective operators at low energies.

\begin{figure}[!t]
\begin{center}
\includegraphics[width=0.48 \textwidth]{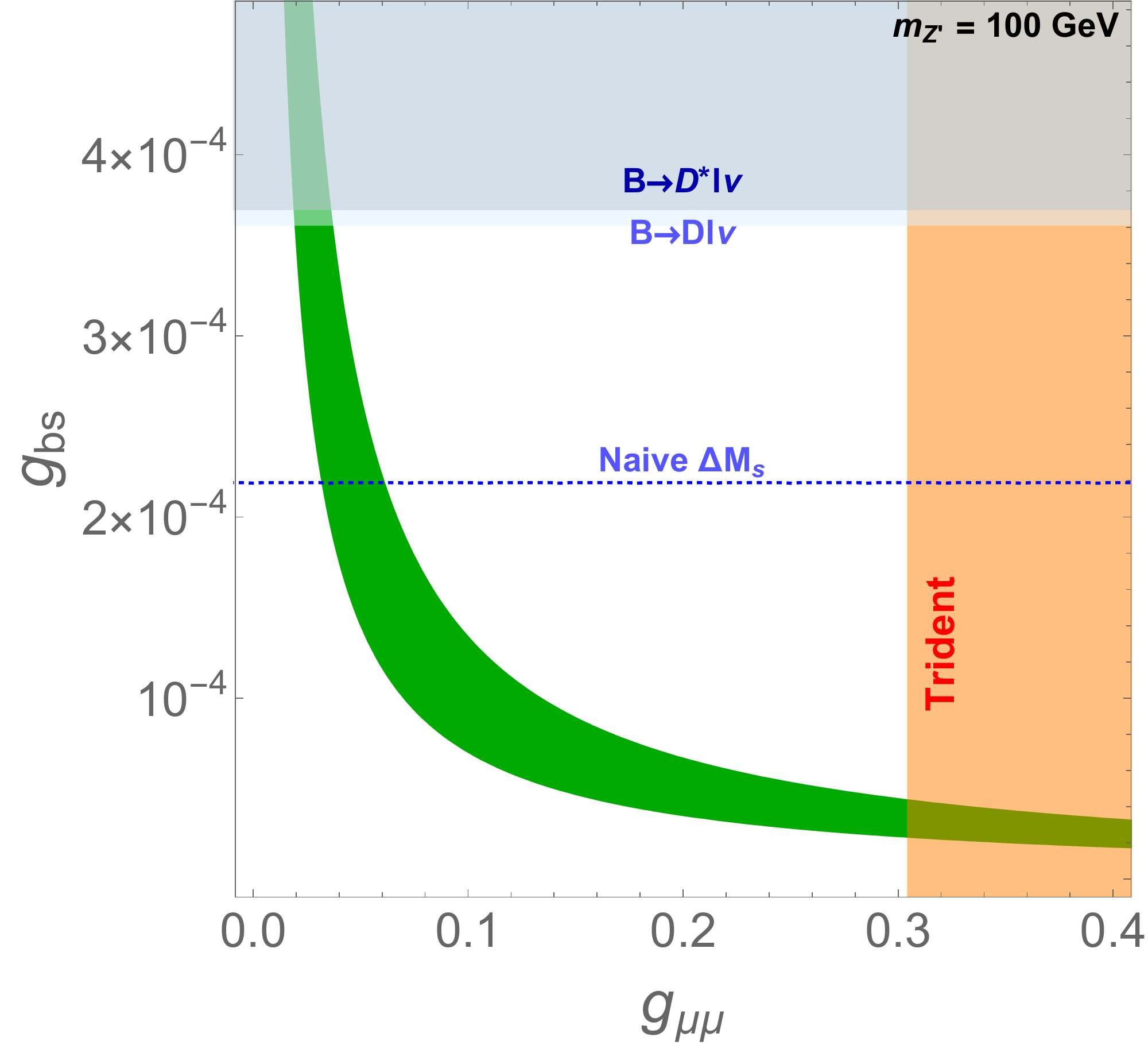}
\quad 
\includegraphics[width=0.48 \textwidth]{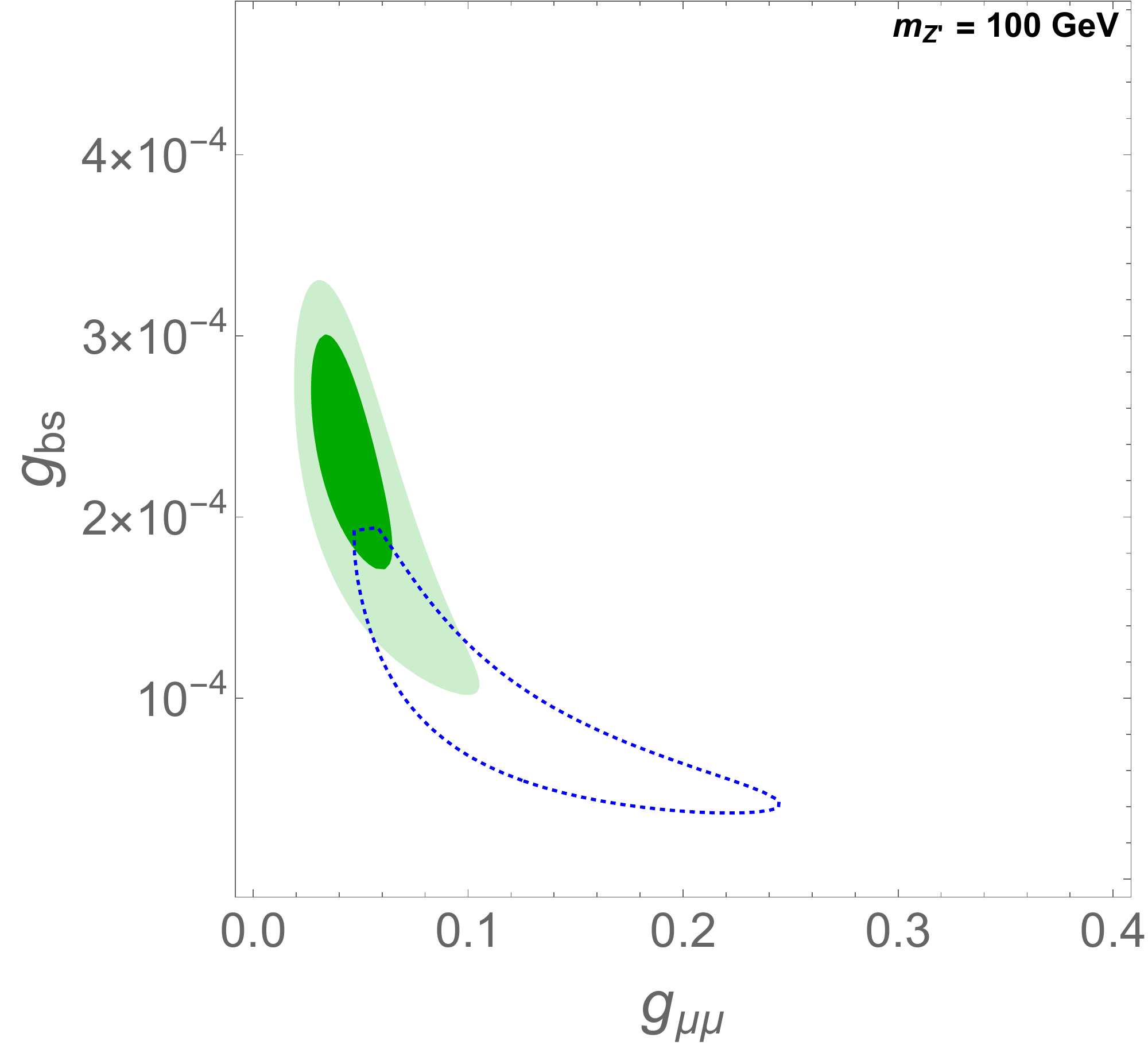}
\caption{
Left: the parameter space in the $(g_{\mu\mu},g_{bs})$ plane for $m_{Z'} = 100 \GeV$ preferred at 68\% CL by the $b\to s\ell\ell$ anomalies (parabolic green band), compared to  the regions excluded at  99\% CL. by trident neutrino production (vertical orange band), and $B \to D^{(*)} \ell \nu$ (horizontal blue bands).  
We also show (dotted blue line) where the naive $\Delta M_s$ constraints would lie as applied e.g. in Ref.~\cite{Falkowski:2018dsl}. 
Right: the 68\% (dark green) and 95\%~CL (light green) regions preferred by the combination of inputs from the $B$-meson anomalies, trident, the $B \to D^{(*)} \ell \nu$ constraints. The dotted blue contour shows the 68\% contour when $B \to D^{(*)} \ell \nu$ constraints are replaced by the naive $\Delta M_s$ ones.
} 
\label{fig:bszprimereloaded}
\end{center}
\end{figure}

\section{Conclusions and outlook}
\label{sec:conclusions}
\setcounter{equation}{0}

In the present article we have discussed the role played by the CKM matrix in the search for NP in the model-independent context of the SMEFT. We recall that the determination of the CKM parameters themselves are then affected by the presence of dimension-6 operators, and that the results from SM global fits combining all available observables performed by CKMfitter~\cite{Charles:2004jd} or UTfit~\cite{Bona:2007vi} cannot be used directly to exploit other flavour constraints involving the CKM matrix.

We have identified a set of four observables: 
\beq
\Gamma(K\to\mu\nu_\mu)/\Gamma(\pi \to\mu\nu_\mu)\ , \quad \Gamma(B\to\tau\nu_\tau)\ , \quad \Delta M_d\ , \quad \Delta M_s \ , 
\eeq 
which are deemed appropriate to determine the CKM parameters in the context of the SMEFT, based on the accuracy of the measurements, the theoretical understanding of their computation, the precision reached on the hadronic inputs, and the simplicity of their calculation
within the SMEFT. We have determined the NP corrections to these four processes 
in the low-energy EFT (below the weak scale), and expressed these corrections in terms of SMEFT contributions by running from the low hadronic scale to the weak scale and performing a matching at the latter.
The corrections from dimension-six operators can then be included in the definition of the ``tilde parameters" $\widetilde{W}_j \equiv \{\tilde\lambda, \tilde A, \tilde \rho, \tilde \eta\}$ (in a procedure similar to one-loop renormalisation), and constraints on $\widetilde{W}_j$ can be extracted. 
Our results for the tilde Wolfenstein parameters are:
\eq{
\tilde\lambda = 0.22537(46)\ , \quad
\tilde A = 0.828(21) \ , \quad
\trho = 0.194(24) \ ,\quad
\teta = 0.391(48) \ .
}
In this exercise, particular attention must be paid to hadronic inputs from lattice QCD simulations: we select data with a lattice scale set by pure QCD quantities (such as bound-state masses) and not by quantities involving the weak interaction (such as decay constants) and potentially modified by NP.

We have also discussed several applications (leptonic meson decays, hadronic $W$ decays, and constraints on a $Z'$ model) to illustrate how our approach leads to a clear interpretation of the measurements of quark-flavour observables and a separation between NP contributions coming from the determination of the CKM parameters and those linked to the process of interest.

The result of our analysis is a set of observables that differs from more usual choices. Indeed, the generality of our approach prevents us from using a rather common option, i.e. using only tree-level processes to extract the CKM parameters. 
It is often advocated that SM loop-level transitions are much more sensitive than SM tree-level transitions to NP effects, so that
tree-level transitions should be used preferentially to determine the CKM parameters. One cannot assume this premise in the general SMEFT set-up. In addition,
a hierarchy of NP effects between SM tree- and loop-level transitions is not supported by the current $B$-anomalies
-- where the (potential) BSM contribution relative to the SM is of the same order in $b\to c\tau\nu$ (SM tree) and $b\to s\mu\mu$ (SM loop) transitions (see e.g.~\cite{Bifani:2018zmi}). It is also 
not present in well-known theoretical frameworks such as Minimal Flavour Violation~\cite{Chivukula:1987py,Hall:1990ac,D'Ambrosio:2002ex}.

The present article outlines the procedure to determine the CKM parameters in SMEFT analyses where only a subset of all flavour observables is taken into account. 
We have proposed a choice of input observables that we consider optimal at this point. This might change in the future if, e.g., new theory developments appear and/or if experimental measurements improve the consistency or the accuracy of some of the other constraints.
Moreover, this choice among observables is not needed in a fit that includes all the measurements that are most sensitive to the CKM parameters, taking into account all correlations. In this case, all observables contribute to the bounds on all the parameters of the fit, and the separation between the processes (mainly) used to extract the CKM parameters and processes (mainly) used to set NP bounds is only useful in order
to illustrate their different sensitivities to each type of parameters. 
This approach was illustrated in studies of NP restricted to $\Delta F=2$ transitions performed by
UTfit~\cite{Bona:2007vi} and CKMfitter~\cite{Charles:2004jd,Lenz:2010gu,Lenz:2012az,Charles:2013aka}.
But such analysis was possible because of the very simple structure of the NP scenario considered. 
In more general settings, and in particular in the full SMEFT case, it is at the moment not possible to proceed in the same way. 
Indeed, in global SMEFT analyses so far the CKM parameters are not treated as free variables, and NP effects affecting their extraction are not taken into account, as discussed in Refs.~\cite{Brivio:2017btx,Aebischer:2018iyb}. 
Our work allows one to overcome this limitation, providing an appropriate framework to consistently include such NP effects as well as the uncertainty on the CKM parameters. 

We have discarded here some of the observables because in the general SMEFT they have complicated expressions involving unknown BSM hadronic matrix elements 
that cannot be easily computed or connected to other hadronic inputs through symmetries.
There are however more specific cases of NP where the expressions of these observables are simple and could provide interesting alternatives to the subset chosen here. The simplest example consists in the case of NP contributions with an SM-like structure for the additional operators so that only $(V-A)\times (V-A)$ charged currents are generated (see, e.g., Ref.~\cite{Brod:2014bfa}).
In practical terms, it would then be useful to include these simplified expressions to be used only if the necessary assumptions are satisfied by the underlying NP operators. One would thus be able to recover in a trivial algorithmic way the setup described above where tree-level processes are approximately free of NP,
or to recover the most accurate results from the SM global fit when the SM case is considered.

In connection with the flavour anomalies currently observed, the constraints on NP coming for flavour physics should be assessed with a particular attention. These deviations  constitute essential probes of the physics at play at energies beyond the current LHC frontier, and one should aim at exploiting current and forthcoming data in global analyses combining large sets of observables from different sectors of particle physics. A careful determination of the CKM parameters in the SMEFT will thus play an important role in this strategy, which should ultimately provide us with new insights in the structure of the physics at higher energies, beyond the Standard Model.

\section*{Acknowledgements}

We are grateful to David Straub, Jason Aebischer, Javier Fuentes-Mart\'in and Christoph Bobeth  for useful comments on the manuscript.
S.D.G. is supported by the European Commission's Horizon 2020 Programme under the Marie Sk\l{}odowska-Curie grant agreements No
690575, No 674896 and No. 692194.
A.F. is supported by the European Commission's Horizon 2020 Programme under the Marie Sk\l{}odowska-Curie grant agreements No 690575 and No 674896. 
M.F. is supported by a grant from the ``Fondazione Della Riccia'', by the MINECO grant FPA2016-76005-C2-1-P and by Maria de Maetzu program grant MDM-2014-0367 of ICCUB and 2017 SGR 929.
M.G-A. is supported by a Marie Sk\l{}odowska-Curie Individual Fellowship of the European Commission’s Horizon 2020 Programme under contract  number  745954  Tau-SYNERGIES.
J.V. acknowledges funding from the European Union's Horizon 2020 research and innovation programme under the Marie Sklodowska-Curie grant agreement No 700525, `NIOBE'.


\appendix
\renewcommand{\theequation}{\Alph{section}.\arabic{equation}} 

\setcounter{equation}{0}

\section{Matching the LEFT to the SMEFT}
\label{app:LEFTSMEFTmatching}

The relevant SMEFT operators in the Warsaw basis~\cite{Grzadkowski:2010es} onto which the LEFT
operators from \Tab{tabops} match are collected in~\Tab{tabSMEFTops}. 
Here, $\varphi$ is a Higgs field, $\ell(e)$ is a left-(right-)handed lepton field, $q$ is a left-handed quark field and $u(d)$ is an up-(down-)type right-handed quark field, with $\{i,j,k,l\}$ family indices, $D_\mu$ is the covariant derivative, and $T^A$ are the Gell-Mann matrices. 

We consider first the matching conditions for the LEFT operators relevant for the semileptonic charged-current transitions collected in the left column in~\Tab{tabops}.
The tree-level matching conditions to the SMEFT at the EW scale are given by~\cite{Cirigliano:2009wk,Aebischer:2017gaw,Jenkins:2017jig}:
\eqa{ \label{eq:semilepWC_SMEFT}
\big[L_{\nu edu}^{V,LL} (\mu_{\rm EW})\big]_{iixk} &=&
-\frac2{v^2}\,V^*_{kx}+2 V^*_{jx} \big[C_{\ell q}^{(3)}\big]_{iijk}-2V^*_{jx}\big[C_{H q}^{(3)}\big]^*_{kj}-2V^*_{kx}\big[C_{H \ell}^{(3)}\big]_{ii}\ ,
\nonumber\\[2mm]
\big[L_{\nu edu}^{S,RR} (\mu_{\rm EW})\big]_{iixk} &=& V^*_{jx}\big[C_{\ell e q u}^{(1)}\big]_{iijk}\ ,\hspace{10mm}
\big[L_{\nu edu}^{V,LR} (\mu_{\rm EW})\big]_{iixk} =  - \big[C_{H ud}\big]^*_{kx} \ ,\nonumber\\[2mm]
\big[L_{\nu edu}^{T,RR} (\mu_{\rm EW})\big]_{iixk} &=& V^*_{jx}\big[C_{\ell e q u}^{(3)}\big]_{iijk}\ ,\hspace{10mm}
\big[L_{\nu edu}^{S,RL} (\mu_{\rm EW})\big]_{iixk} = \big[C_{\ell e dq}\big]_{iixk}\ ,
}
where we assume all SMEFT parameters given at the EW scale by default. 
These matching conditions refer to the coefficients of the LEFT operators in the mass basis (where $x=\{d,s,b\}$, $i=\{e,\mu,\tau\}$ and $k=\{u,c\}$),
which are related to the ones in the weak basis by the CKM matrix,
by virtue of~\Eq{ditodx}.

The LEFT operators relevant for $\bar B_q- B_q$ mixing are collected in the second column of~\Tab{tabops}.
The tree-level matching to the SMEFT at order $1/\Lambda^2$ is given by~\cite{Jenkins:2017jig}:
\eqa{ \label{eq:neutralWC_SMEFT}
\big[L_{dd}^{V,LL}(\mu_{EW})\big]_{dbdb}  &=&
\big[L_{dd}^{V,LL}(\mu_{\rm EW})\big]_{dbdb}^{\rm SM} + V_{id} V_{jb}^* V_{kd} V_{\ell b}^* \,\Big( \big[C_{qq}^{(1)}\big]_{ijkl}+\big[C_{qq}^{(3)}\big]_{ijkl}\Big) \ ,
\nonumber\\[2mm]
\big[L_{dd}^{V,RR}(\mu_{EW})\big]_{dbdb}  &=& \big[C_{dd}\big]_{1313} \ ,
\nonumber\\[2mm]
\big[L_{dd}^{V1,LR}(\mu_{EW})\big]_{dbdb}  &=& V_{id} V_{jb}^* \,\big[C_{qd}^{(1)}\big]_{ij13} \ ,
\quad
\big[L_{dd}^{V8,LR}(\mu_{EW})\big]_{dbdb}  = V_{id} V_{jb}^* \,\big[C_{qd}^{(8)}\big]_{ij13} \ ,
\nonumber\\[2mm]
\big[L_{dd}^{S1,RR}(\mu_{EW})\big]_{dbdb}  &=& 
\big[L_{dd}^{S1,RR}(\mu_{EW})\big]_{bdbd} =
\big[L_{dd}^{S8,RR}(\mu_{EW})\big]_{dbdb}  =
\big[L_{dd}^{S8,RR}(\mu_{EW})\big]_{bdbd} = 0,  \nnl
}
and analogously for $d\to s$. Some $1/\Lambda^4$ terms are known~\cite{Aebischer:2015fzz,Jenkins:2017jig}, but have been dropped here.
In addition, one-loop matching corrections give non-zero contributions to $L_{dd}^{S1,RR}$ and $L_{dd}^{S8,RR}$, which we have
also ignored, for simplicity. Some of these can be found in~\Ref{Aebischer:2015fzz}.

In the SM limit only the operator $L_{dd}^{V,LL}$ is non-zero, starting at one loop. The two-loop result~\cite{Buras:1990fn} is given by (for $q=\{d,s\}$):
\eq{ \label{eq:C1_SM}
C_{1,\rm SM}^{(q)} \equiv \big[L_{dd}^{V,LL}(\mu_{\rm EW})\big]_{qbqb}^{\rm SM} = - \frac{M_W^2}{32\pi^2 v^4} (V_{tq} V_{tb}^*)^2\, S_1(\mu_{\rm EW})\ ,
}
where\footnote{
The function $S_1$ evaluated at the low scale $\mu_b$ is $S_1(\mu_b) = U_{11}(\mu_{\rm EW},\mu_b)\,S_1(\mu_{\rm EW}) \simeq 0.8583 \times 2.3124 = 1.985$, and corresponds to what is traditionally denoted by $\hat \eta_B\,S_0(x_t)\simeq 0.83798 \times 2.36853 = 1.985$,
where the factor $\hat \eta_B$ encodes NLO QCD matching corrections and running effects simultaneously
(see e.g. Ref.~\cite{DiLuzio:2017fdq}). In this way we have made the separation between EW matching and RGE running explicit.
Note that in our conventions the sign of the Wilson coefficients $C_i$ is opposite than that in~\Ref{Gabbiani:1996hi}. 
}
$S_1(\mu_{\rm EW}) \simeq 2.3124$ contains the NLO (two-loop) QCD correction to the SM matching~\cite{Buras:1990fn} at the matching scale $\mu_{\rm EW}=M_Z$ (correcting the well-known one-loop result $S_0(x_t)\simeq 2.369$, where $S_0(x)= (x^4 - 12x^3 + 15x^2 - 4x + 6x^3 \ln{x})/4(x-1)^3$ is the Inami-Lin function~\cite{Inami:1980fz}).

In the computation of the mass differences we will use the traditional ``SUSY basis" for the $\Delta F=2$ operators~\cite{Gabbiani:1996hi},
for which the matrix elements are explicitly known.
These are denoted by  $O^{(q)}_{1,\dots,5}$, $\widetilde O^{(q)}_{1,2,3}$,
and their relation to the LEFT basis (in $d=4$) is given by
\begin{align}
&\big[\cO_{dd}^{V,LL}\big]_{qbqb} = \cO_1^{(q)} \ ,            &&   \big[\cO_{dd}^{V,RR}\big]_{qbqb} = \widetilde \cO_1^{(q)}\ , \nonumber\\
&\big[\cO_{dd}^{V1,LR}\big]_{qbqb} = -2\,\cO_5^{(q)}\ ,        &&   \big[\cO_{dd}^{V8,LR}\big]_{qbqb}   = -\cO_4^{(q)} + \cO_5^{(q)}/N_c\ , \nonumber\\
&\big[\cO_{dd}^{S1,RR}\big]_{qbqb} = \widetilde \cO_2^{(q)}\ , &&   \big[\cO_{dd}^{S1,RR}\big]_{bqbq}^\dagger  = \cO_2^{(q)}\ ,  \nonumber\\
&\big[\cO_{dd}^{S8,RR}\big]_{qbqb} = -\widetilde \cO_2^{(q)}/(2N_c) + \widetilde \cO_3^{(q)}/2\ , 
&& \big[\cO_{dd}^{S8,RR}\big]_{bqbq}^\dagger  = -\cO_2^{(q)}/(2N_c) +  \cO_3^{(q)}/2\ .  
\end{align}

\begin{table}
\centering
\setlength{\tabcolsep}{22pt}
\renewcommand{\arraystretch}{1.7}
\begin{tabular}{@{}ll@{}}
\toprule
Semileptonic  & $\mu$ decay \\ 
\hline
$\big[Q_{H \ell}^{(3)}\big]_{ij} = (\varphi^\dagger i \overleftrightarrow{D}_\mu^I \varphi)(\bar\ell_i \sigma^I \gamma^\mu \ell_j)$
&
$\big[Q_{\ell\ell}\big]_{ijkl} = (\bar{\ell}_i\gamma^\mu \ell_j)(\bar{\ell}_k\gamma_\mu \ell_l)$
\\[1.5mm]
\cmidrule(lr){2-2}
$\big[Q_{H q}^{(3)}\big]_{ij} = (\varphi^\dagger i \overleftrightarrow{D}_\mu^I \varphi)(\bar q_i \sigma^I \gamma^\mu q_j)$
&
$\Delta F=2$\\
\cmidrule(lr){2-2}
$\big[Q_{H ud}\big]_{ij} = i(\widetilde \varphi^\dagger D_\mu \varphi)(\bar u_i \gamma^\mu d_j) + h.c.$
&
$\big[Q_{qq}^{(1)}\big]_{ijkl} = (\bar{q}_i\gamma^\mu q_j)(\bar{q}_k\gamma_\mu q_l)$
\\
$\big[Q_{\ell q}^{(3)}\big]_{ijkl} = (\bar\ell_i\gamma^\mu \sigma^I \ell_j)(\bar{q}_k\gamma_\mu \sigma^I q_l)$
&
$\big[Q_{qq}^{(3)}\big]_{ijkl} = (\bar{q}_i\gamma^\mu \sigma^I q_j)(\bar{q}_k\gamma_\mu \sigma^I q_l)$
\\
$\big[Q_{\ell e q u}^{(1)}\big]_{ijkl} = (\bar{\ell}_i^m e_j)\epsilon_{mn}(\bar{q}_k^n u_l)$
&
$\big[Q_{dd}\big]_{ijkl}  = (\bar{d}_i\gamma^\mu d_j)(\bar{d}_k\gamma_\mu d_l)$
\\
$\big[Q_{\ell e q u}^{(3)}\big]_{ijkl} = (\bar{\ell}_i^m \sigma_{\mu\nu} e_j)\epsilon_{mn}(\bar{q}_k^n \sigma^{\mu\nu} u_l)$
&
$\big[Q_{qd}^{(1)}\big]_{ijkl}  = (\bar{q}_i\gamma^\mu q_j)(\bar{d}_k\gamma_\mu d_l)$
\\
$\big[Q_{\ell e d q}\big]_{ijkl} = (\bar{l}_i e_j)(\bar{d}_k q_l)$
&
$\big[Q_{qd}^{(8)}\big]_{ijkl}  = (\bar{q}_i\gamma^\mu T^A q_j)(\bar{d}_k\gamma_\mu T^A d_l)$
\\[1mm]
\bottomrule
\end{tabular}
\caption{Operators in the SMEFT relevant for $\mu$ decay, semileptonic and $\Delta F=2$ transitions.}
\label{tabSMEFTops}
\end{table}

\setcounter{equation}{0}

\section{Renormalisation Group Evolution}
\label{app:LEFTrunning}

We start with the leptonic decays $P\to \ell \bar\nu_\ell$ discussed in~\Sec{sec:Pell2}.
Using three-loop QCD running plus one-loop QED running ~\cite{Aebischer:2017gaw,Gonzalez-Alonso:2017iyc}, 
the parameters $\epsilon_X^{lxy}$ defined in \eref{epsilondef} are given by
\small
\eqa{
\label{eq:semilepWC_LEFT}
\epsilon_A^{\mu ud} &=& - 1.0094 
- \frac{v^2}{2 V_{ud}} \Big(
1.0094\, \big[L_{\nu edu}^{V,LL} (\mu_{\rm EW})\big]^*_{\mu\mu d u}
-1.0047\, \big[L_{\nu edu}^{V,LR} (\mu_{\rm EW})\big]^*_{\mu\mu d u}
\Big)\ ,
\nonumber\\[2mm]
\epsilon_P^{\mu ud} &=& 
- \frac{v^2}{2 V_{ud}} \Big(
1.73  \big[L_{\nu edu}^{S,RR} (\mu_{\rm EW})\big]^*_{\mu\mu d u}
- 1.73  \big[L_{\nu edu}^{S,RL} (\mu_{\rm EW})\big]^*_{\mu\mu d u}
- 0.024  \big[L_{\nu edu}^{T,RR} (\mu_{\rm EW})\big]^*_{\mu\mu d u}
\Big)\ ,
\nonumber\\[4mm]
\epsilon_A^{\mu us} &=& - 1.0094 
- \frac{v^2}{2 V_{us}} \Big(
1.0094\, \big[L_{\nu edu}^{V,LL} (\mu_{\rm EW})\big]^*_{\mu\mu s u}
-1.0047\, \big[L_{\nu edu}^{V,LR} (\mu_{\rm EW})\big]^*_{\mu\mu s u}
\Big)\ ,
\nonumber\\[2mm]
\epsilon_P^{\mu us} &=& 
- \frac{v^2}{2 V_{us}} \Big(
1.73  \big[L_{\nu edu}^{S,RR} (\mu_{\rm EW})\big]^*_{\mu\mu s u}
- 1.73  \big[L_{\nu edu}^{S,RL} (\mu_{\rm EW})\big]^*_{\mu\mu s u}
- 0.024  \big[L_{\nu edu}^{T,RR} (\mu_{\rm EW})\big]^*_{\mu\mu s u}
\Big)\ ,
\nonumber\\[4mm]
\epsilon_A^{\tau ub} &=& - 1.0075 
- \frac{v^2}{2 V_{ub}} \Big(
1.0075\, \big[L_{\nu edu}^{V,LL} (\mu_{\rm EW})\big]^*_{\tau\tau b u}
-1.0038\, \big[L_{\nu edu}^{V,LR} (\mu_{\rm EW})\big]^*_{\tau\tau b u}
\Big)\ ,
\\[2mm]
\epsilon_P^{\tau ub} &=& 
- \frac{v^2}{2 V_{ub}} \Big(
1.45  \big[L_{\nu edu}^{S,RR} (\mu_{\rm EW})\big]^*_{\tau\tau b u}
- 1.45  \big[L_{\nu edu}^{S,RL} (\mu_{\rm EW})\big]^*_{\tau\tau b u}
- 0.018  \big[L_{\nu edu}^{T,RR} (\mu_{\rm EW})\big]^*_{\tau\tau b u}
\Big)\ ,
\nonumber
}
in terms of the LEFT Wilson coefficients at the EW scale.
These are, in turn, related to the SMEFT coefficients via~\Eq{eq:semilepWC_SMEFT}.

To describe $\bar B_q- B_q$ mixing we  use the Wilson coefficients in the SUSY basis (see \eref{neutralWC_SMEFT}) in the $\overline{\rm MS}$ scheme of~\Ref{Buras:2000if}, at the scale $\mu_b=4.3\GeV$, in accordance with the matrix elements of the operators provided in~\cite{Carrasco:2013zta}.
We call these Wilson coefficients $C^{(q)}_{1,\dots,5}$, $\widetilde C^{(q)}_{1,2,3}$. In order to relate these coefficients to the ones
at the EW scale, we use the NLO evolution matrix given in~\Ref{Buras:2000if}  (in the same scheme but different basis -- see Section~3 of~\Ref{Mescia:2012fg} for detail), running $\alpha_s$ at four loops~\cite{Chetyrkin:2000yt}.
In the end we find, for $\mu_{\rm EW} = M_Z$ and $q=\{d,s\}$: 
\eqa{ \label{eq:neutralWC_LEFT}
C^{(q)}_1 &=& 0.858\,\big[L_{dd}^{V,LL}(\mu_{EW})\big]_{qbqb} \ ,\nonumber\\[2mm]
C^{(q)}_2 &=& 1.545\,\big[L_{dd}^{S1,RR}(\mu_{EW})\big]_{bqbq}^*  -0.387\,\big[L_{dd}^{S8,RR}(\mu_{EW})\big]_{bqbq}^* \ ,\nonumber\\[2mm]
C^{(q)}_3 &=& -0.047\,\big[L_{dd}^{S1,RR}(\mu_{EW})\big]_{bqbq}^*  +0.312\,\big[L_{dd}^{S8,RR}(\mu_{EW})\big]_{bqbq}^* \ ,\nonumber\\[2mm]
C^{(q)}_4 &=& -0.755\,\big[L_{dd}^{V1,LR}(\mu_{EW})\big]_{qbqb}   -1.940\,\big[L_{dd}^{V8,LR}(\mu_{EW})\big]_{qbqb} \ ,\nonumber\\[2mm]
C^{(q)}_5 &=&  -1.856\,\big[L_{dd}^{V1,LR}(\mu_{EW})\big]_{qbqb}  +0.237\,\big[L_{dd}^{V8,LR}(\mu_{EW})\big]_{qbqb} \ ,\nonumber\\[2mm]
\widetilde C^{(q)}_1 &=& 0.858\,\big[L_{dd}^{V,RR}(\mu_{EW})\big]_{qbqb} \ ,\nonumber\\[2mm]
\widetilde C^{(q)}_2 &=& 1.545\,\big[L_{dd}^{S1,RR}(\mu_{EW})\big]_{qbqb}  -0.387\,\big[L_{dd}^{S8,RR}(\mu_{EW})\big]_{qbqb} \ ,\nonumber\\[2mm]
\widetilde C^{(q)}_3 &=& -0.047\,\big[L_{dd}^{S1,RR}(\mu_{EW})\big]_{qbqb}  +0.312\,\big[L_{dd}^{S8,RR}(\mu_{EW})\big]_{qbqb} \ .
}

\setcounter{equation}{0}

\section{NP shifts to the CKM parameters beyond linear order}
\label{app:quadratic}

The relation between the NP shifts in the Wolfenstein parameters, $\delta W = \{ \delta \lambda, \delta A, \delta \bareta, \delta \barrho \}$, and the NP contributions to the chosen observables, denoted by $\Delta\equiv \{ \Delta_{K/\pi},\Delta_{B\tau2}, \Delta_{\Delta M_{d}}$, $\Delta_{\Delta M_{s}} \}$, 
is in general a complicated non-linear equation. However, assuming that NP is a small perturbation one can expand that equation around the SM point to any given order, and obtain its unique solution. 
Let us note that, by construction, this approach discards possible additional solutions where the NP correction is comparable to or larger than the SM contribution. 

The generalization of \eref{deltaWmatricial} to include quadratic NP terms is the following
\bea
\label{eq:quadratic}
\delta W = \left( 1 - M \Delta' + F \right) M \Delta - M \Delta^2~,
\eea
where the NP corrections to the observables, $\Delta$, should include quadratic corrections. We have introduced the quantities
\eq{
\Delta'_{ij} \equiv \partial \Delta_i / \partial W_j,
\;
\Delta^2 \equiv \big\{ \Delta_{K/\pi}^2,\Delta_{B\tau2}^2, \Delta_{\Delta M_{d}}^2,\Delta_{\Delta M_{s}}^2 \big\},
\;
M \equiv \left({\cal O}'\right)^{-1} {\cal O},
\;
F \equiv \left({\cal O}'\right)^{-1} P,
}
where $M$ corresponds to the matrix in \eref{deltaWmatrix}, and
\eq{
{\cal O} = {\rm diag} \big(\Gamma(K\to \mu \nu)/\Gamma(\pi \to \mu\nu), \Gamma(B\to\tau\nu), \Delta M_d, \Delta M_s \big)_{\rm SM}
}
are the SM expressions of the input observables.
The matrices ${\cal O}'$ and $P$ are defined by
\bea
{\cal O}'_{ij} &\equiv& \partial {\cal O}_{ii} / \partial W_j~,\\
 P_{ij} &\equiv&  \frac{1}{2} \frac{\partial^2 {\cal O}_{ii}}{ \partial W_j \partial W_k} \left( M \Delta_{\rm NP} \right)_k~\,. 
\eea
All quantities above are implicitly evaluated at $W = \widetilde W$.


\bibliographystyle{./JHEP.bst} 
\bibliography{ckmpp}

\end{document}

%% file: shortcuts.tex
\newcommand{\fref}[1]{Fig.~\ref{fig:#1}} 
\newcommand{\eref}[1]{Eq.~\eqref{eq:#1}}

\newcommand{\sref}[1]{Section~\ref{sec:#1}}
\newcommand{\cref}[1]{Chapter~\ref{ch:#1}}

\newcommand{\nnl}{\nonumber \\}

\newcommand{\beq}{\begin{equation}} 
\newcommand{\eeq}{\end{equation}} 
\newcommand{\ba}{\begin{array}}  
\newcommand{\ea}{\end{array}} 
\newcommand{\bea}{\begin{eqnarray}}  
\newcommand{\eea}{\end{eqnarray} }  
\newcommand{\be}{\begin{eqnarray}}  
\newcommand{\ee}{\end{eqnarray} }  
\newcommand{\bal}{\begin{align}}
\newcommand{\eal}{\end{align}}   
\newcommand{\bi}{\begin{itemize}}  
\newcommand{\ei}{\end{itemize}}  
\newcommand{\ben}{\begin{enumerate}}  
\newcommand{\een}{\end{enumerate}}  
\newcommand{\bc}{\begin{center}}
\newcommand{\ec}{\end{center}} 
\newcommand{\bt}{\begin{table}}
\newcommand{\et}{\end{table}}  
\newcommand{\btb}{\begin{tabular}}
\newcommand{\etb}{\end{tabular}}  
\newcommand{\bvec}{\left ( \ba{c}}
\newcommand{\evec}{\ea \right )}

\newcommand{\cO}{{\mathcal O}} 
\newcommand{\co}{{\mathcal O}} 
\newcommand{\cL}{{\mathcal L}}

\newcommand{\gev}{\mathrm{GeV}}

\newcommand{\re}{{\mathrm{Re}} \,}

\newcommand{\tlambda}{\tilde \lambda}  
\newcommand{\tA}{\tilde A}
\newcommand{\teta}{\tilde \eta}  
\newcommand{\trho}{\tilde \rho}  
\def\hc{{\rm h.c.}} 

\newcommand{\barrho}{ \bar{\rho}}
\newcommand{\bareta}{ \bar{\eta}}